\documentstyle[aps]{revtex}

\begin{document} 
\draft
\preprint{RU9618}
\title{Effects of the Spin-Orbit and Tensor Interactions on the
$M1$ and $E2$ Excitations in Light Nuclei}
\author{M. S. Fayache$^{1,3}$, Y.Y. Sharon$^1$ and
L. Zamick$^{1,2}$, P. von Neumann-Cosel$^2$ and A. Richter$^2$} 
\address{(1) Department of Physics and Astronomy, Rutgers University,
Piscataway, New Jersey 08855\\
(2) Institut fur Kernphysik, Technische Hochschule, Darmstadt,
D-64289 Darmstadt, Germany\\
(3) D\'{e}partement de Physique, Facult\'{e} des Sciences de Tunis,
Tunis 1060, Tunisia} 
\date{\today}
\maketitle
\begin{abstract}
The effects of varying the spin-orbit and tensor components of a 
realistic interaction on $M1$ excitation rates and $B(E2)'s$ are
studied on nuclei in the $0p$ and $1s-0d$ shells. Not only the total
$M1$ but also the spin and orbital parts separately are studied. 
The single-particle energies are first calculated with the same
interaction that is used between the valence nucleons. Later this
stringent condition is relaxed somewhat and the $1s$ level is raised
relative to $0d$. For nuclei up to $^{28}Si$, much better results i.e
stronger $B(M1)$ rates are obtained by increasing the strength of the
spin-orbit interaction relative to the free value. This is probably
also true for $^{32}S$, but $^{36}Ar$ presents some difficulties. 
The effects of weakening the tensor interaction are also studied. 
On a more subtle level, the optimum spin-orbit interaction in the
lower half of the $s-d$ shell, as far as $M1$ excitations are
concerned, is substantially larger than the difference
$E(J=3/2^+)_1-E(J=5/2^+)_1=5.2~MeV$ in $^{17}O$. A larger spin-orbit
splitting is also needed to destroy the triaxiality in $^{22}Ne$. 
Also studied are how much $M1$ orbital and spin strength lies in an
observable region and how much is buried in the grass at higher
energies. It is noted that for many nuclei the sum
$B(M1)_{orbital}+B(M1)_{spin}$ is very close to $B(M1)_{total}$,
indicating that the summed cross terms are very small.  
\end{abstract}

\section{Introduction}
Our purpose is to see how various components of the realistic
nucleon-nucleon interaction affect nuclear properties. In particular,
we shall investigate magnetic dipole excitations to $J=1^+$ states of
even-even nuclei -these are spin and orbital modes, and as a counter
point we shall look at low lying $2^+$ states which for deformed open
shell nuclei have very little spin content. We will examine selected
nuclei in the $0p$ and $1s-0d$ shell. 

At first thought it may seem that such an investigation is unnecessary
because excellent fits have been obtained by Cohen and
Kurath~\cite{ck} in the $p$ shell and by Wildenthal and
Brown~\cite{bw1} in 
the $1s-0d$ shell. These works are of extreme importance if for no
other reason than that they showed that the shell model works well
beyond many peoples expectations -well enough in fact so that the
authors in collaboration with the experimental groups could
meaningfully go one step further and discuss effects of exchange
currents on $M1$'s in a systematic way~\cite{rw,bw2}. The works
brought order out of the myriad of energy levels that were discussed
by experimentalists. 

However, the truly excellent fits were obtained with empirical
effective interactions e.g. the Wildenthal interaction in the $sd$
shell~\cite{wil}. Such an interaction implicitly has in it several effects
e.g. core polarization, relativistic phenomenology etc. It is
impossible to sort out these effects from an empirical interaction
expressed as hundreds of matrix elements. 

Our intention is to start with a realistic interaction which will
admittedly not give as good a fit to the data as the above mentioned
empirical interactions. But the discrepancies will show what is
missing and what has to be done. The interaction that we use -the
($x,y$) interaction- has been discussed elsewhere~\cite{zz}. Suffice
it to say that it is written as follows:

\[V(r)=V_c(r)+xV_{s.o.}+yV_t\]

\noindent where $s.o.$ stands for the two-body spin-orbit interaction and $t$
for the tensor interaction and $V_c(r)$ is everything else especially
the central interaction. For $x=1,~y=1$, the interaction gives a good
fit to the Bonn A matrix elements~\cite{zz}. We can study the effects of
changing the spin-orbit and tensor interactions by
varying $x$ and $y$. 

Most calculations which claim to use realistic interactions are really
doing hybrid calculations in which the residual two-body matrix
elements are carefully calculated but the single-particle energies are
obtained from experiment. Here on the other hand we shall use the same
interaction to {\em calculate} the single-particle matrix elements as
was used to calculate the residual interaction. We feel this is the
only way to {\em really} find out the effects of such an
interaction. However, to overcome the deficiencies of using harmonic
oscillator wavefunctions, we will also consider one case in which the
relative spacing of the $1s$ and $0d$ single-particle energies are
changed (likewise for $p$ and $0f$).

We shall perform calculations with four sets of ($x,y$): (1,1),
(1.5,1), (1.,0.5) and (1.5,0.5). We are motivated in these choices by
the ideas presented by many that the spin-orbit interaction inside a
nucleus should be stronger than what is deduced form nucleon-nucleon
scattering and the tensor interaction weaker. We shall also consider
one more case in which, for the $1s-0d$ shell, the $1s$
single-particle energy is moved up by 2 $MeV$ relative to the $0d$
centroid ($\Delta \epsilon_s=2~MeV$). This is to overcome the defects
of using harmonic oscillator wavefunctions, and is equivalent to the
Nilsson prescription of adding a $D\vec{l}\cdot \vec{l}$ term to the
Hamiltonian.  

A stronger spin-orbit interaction arises for example in the Walecka
model, also known as Dirac phenomenology~\cite{wal}. There is a
parameter in the theory called the Dirac effective mass which arises
from sigma exchange. The ratio $m_D/m$ is less than one and the
spin-orbit interaction is inversely proportional to $m_D/m$. In
another formulation of Wiring et. al., the stronger spin-orbit
interaction comes from a three-body interaction~\cite{wir}. 

Concerning the weakening of the tensor interaction inside a nucleus,
two basic ideas have been advanced. One is the universal scaling
argument by G.E. Brown and collaborators~\cite{brown}. The basic idea
is that all mesons inside the nucleus except for the pion become less
massive. Hence the range of the nucleon-nucleon interaction due to a
given boson exchange becomes larger. In particular, the $\rho$ meson
exchange will be of larger range. This gives a larger repulsive
contribution to the tensor interaction which will cancel the
attraction due to the pion. The increased cancellation will result in
a weaker tensor interaction.

On the other hand Fayache, Zheng and Zamick have advocated a
self-weakening mechanism for the tensor interaction~\cite{zzf}. They show that
the effects of higher-shell admixtures makes the tensor interaction in
a $\Delta N=0$ (one major shell) space appear weaker. Thus, at least
for {\em nuclear structure calculations}, no mechanism beyond higher
shell admixtures seems to be needed. 

However, the theory of Brown et. al.~\cite{brown} also pertains to
nucleon-nucleus scattering problems at intermediate to high
energies. Here the anomalies for the $q^2$ behaviour of several
Wolfenstein parameters can be explained by the universal scaling
arguments, but it has not been demonstrated that better nuclear
structure calculations would also resolve these anomalies. 

\section{The Magnetic Dipole Operator}

In what follows we discuss present calculations for what we call
$B(M1)_{physical}$, $B(M1)_{orbital}$ and $B(M1)_{spin}$. In terms of
a magnetic dipole operator, the $B(M1)$ is defined as 

\[\frac{1}{(2J_i+1)}\sum_{M_f,\mu, M_i}|\langle
\Psi_{M_f}^{J_f}~A~_{\mu}~\Psi_{M_i}^{J_i}\rangle |^2 \]

\noindent where 

\[\vec{A}=\sqrt{3/4\pi}\left [
\sum_{protons}(g_{l\pi}\vec{l_i}+g_{s\pi}\vec{s_i}) +
\sum_{neutrons}(g_{l\nu}\vec{l_i}+g_{s\nu}\vec{s_i})\right]\]

For the three cases considered, the following parameters were used:

\begin{tabbing}
Physical~~~~\=$g_{l\pi}=1$~~~~\=$g_{s\pi}=5.586$~~~~\=$g_{l\nu}=0~~g_{s\nu}=
-3.826$\\
Orbital \> $g_{l\pi}=1$ \> $g_{s\pi}=0$ \> $g_{l\nu}=0~~g_{s\nu}=0$\\
Spin \> $g_{l\pi}=0$ \>$g_{s\pi}=5.586$~~~~\=$g_{l\nu}=0~~g_{s\nu}=
-3.826$\\
\end{tabbing}

In other words, for the $B(M1)_{orbital}$ we set the spin $g$ factors
to zero and for $B(M1)_{spin}$ we set the orbital $g$ factors to
zero. 

One can also write the magnetic moment operator as
$\vec{A}=\vec{A}_{scalar} + \vec{A}_{vector}\tau_z$ where $\tau_z=+1$
for the proton and -1 for the neutron. For the physical operator, 

\[g_s(scalar)=\frac{g_{s\pi}+g_{s\nu}}{2}=0.88\] 

\noindent and

\[g_s(vector)=\frac{g_{s\pi}-g_{s\nu}}{2}=4.706\]  

Both $g_l(scalar)$ and $g_l(vector)$ are equal to 0.5. Because
$g_s(vector)$ is much larger than $g_s(scalar)$, $T=0$ to $T=1$
transitions in $N=Z$ nuclei tend to be much larger that
$T=0~\rightarrow T=0$ transitions. The ratio
$(\frac{g_s(vector)}{g_s(scalar)})^2$ is equal to 81. 

There are some interesting limiting cases for $M1$ matrix
elements. For example, in the $SU(4)$ limit, all spin matrix
transition matrix elements from the $J=0^+$ ground state (both
isoscalar and isovector) will vanish. The isoscalar orbital transition
will also vanish. All that is left is the isovector orbital
excitations known as scissors modes. 

For strongly deformed nuclei one approaches this limit. The nuclei in
the lower part of the $s-d$ shell e.g. $^{20}Ne$ and $^{24}Mg$ are
strongly deformed and so we should look for such selection rules
there. However, the spin contributions will not be negligible because
$\frac{g_s(vector)}{g_l(vector)}=9.41$, and even if the spin matrix
elements are strongly suppressed they start out with a big
advantage. 

\section{The Single-Particle Energies}

As mentioned in the introduction, we do not perform a hybrid
calculation. We first obtain the single-particle energies with the same
interaction that is used for the two-particle matrix elements. More to
the point, in our shell model, we perform a calculation for say
$^{20}Ne$ not merely as four nucleons in the $1s-0d$ shell but rather
as twenty nucleons  with sixteen in the closed $0s$ and $0p$
shells. This allows for valence-core interactions which implicitly
generate the single-particle energies. (In a later section we relax
this stringent condition by moving the $1s$ single-particle level
relative to $0d$ in order to overcome the defects of using harmonic
oscillator wavefunctions). 

In Table~I we present the single-particle energies obtained for the
($x,y$) interaction with $x=1,~y=1$ in both a small space and a large
space. By the former we mean that the $^{16}O$ core is inert. In the
large space, we allow $2 \hbar \omega$ excitations in the shell model 
diagonalization and we set the lowest state to zero~\cite{oxbash}. 

We see one defect in that the lower $l$ states come down too low
relative to the higher $l$ states. Whereas in the calculation the
$1s_{1/2}$ comes below $0d_{5/2}$, for $^{17}O$ we know it's the other
way around. Likewise the $1p_{3/2}$ comes 0.94 $MeV$ below $0f_{7/2}$
whereas experimentally the $1p_{3/2}$ is about 2 $MeV$ above
$0f_{7/2}$. This tendency of the low $l$ to come too low has the
effect of making nuclei more deformed than they are experimentally. 

The calculated spin-orbit splitting for $^{15}O$ in the small space is 5.07
$MeV$, somewhat smaller than the splitting between the $J=3/2^-$
excited state of $^{15}O$ and the $J=1/2^-$ ground state. The latter
is 6.1 $MeV$. For $^{17}O$ the splitting $J=3/2^-~-~J=1/2^-$ is 5.562
$MeV$ in the calculation, which compares well with the empirical
splitting of 5.2 $MeV$. 

Going to the large space does not increase the splitting
significantly. For $A=15$ the value is 5.679 $MeV$ while for $A=17$ it
is 5.662 $MeV$. 

An interesting question to ask is whether the empirical procedure
above of identifying the spin-orbit splitting in say $^{17}O$ with
$E(J=3/2^-)-E(J=5/2^-)$ is valid. We shall discuss this in the context
of the systematics of the magnetic dipole transitions that we
calculate. 

\section{Results for $M1$ and $E2$ Excitation Rates as a Function of
$x$ and $y$}

{\bf (a) The Effects of the Spin-Orbit and Tensor Interactions on the
Spin $B(M1)$}

For all nuclei considered, the effect of increasing the spin-orbit
strength parameter $x$ from 1 to 1.5 causes the spin $B(M1)$ (also
denoted by $B_\sigma$ -last column of Table~II) to increase. In
$^{12}C$ the increase (for $y=1$) is from 0.91 to 2.07; in $^{20}Ne$
from 0.59 to 1.28; in $^{24}Mg$ from 1.73 to 3.35, in $^{32}S$ from
2.72 to 8.52; and in $^{36}Ar$ from 2.25 to 5.20. The change is especially large in $^{32}S$. 

The increase in $B_\sigma$ can be understood by noting that this
quantity increases as we go from the $LS$ limit to the $j-j$
limit. Indeed, in the $SU(4)$ limit, the spin $B(M1)$ will vanish. 

In all cases considered, decreasing the tensor interaction also causes
the spin $B(M1)$ to increase. This is consistent with the old
observation of C.W. Wong~\cite{wong} that the tensor interaction in an
open shell 
nucleus acts to some extent like a spin-orbit interaction of the
opposite sign as the main spin-orbit term. Thus, in $^{24}Mg$, when we
go from $x=1~y=1$ to $x=1~y=0.5$ the $B(M1)_{spin}$ (in units of
$\mu_N^2$) goes from 1.73 to 1.94; and in going from $x=1.5~y=1$ to
$x=1.5~y=0.5$, we find that $B(M1)_{spin}$ increases from 3.35 to
4.04. There is a similar behaviour for all other nuclei
considered. The effect of varying the tensor strength $y$ is not quite
as dramatic as that of varying the spin-orbit strength $x$. 

It should be emphasized that there are basic differences between the
tensor and the spin-orbit interactions. For a closed $LS$ plus or
minus one nucleon, the tensor interaction gives no contribution to the
single-particle energy in first order whilst the two-body spin orbit
interaction does. The latter gives the majority amount of the
spin-orbit splitting e.g. of $d_{3/2}$ and $d_{5/2}$ in $^{17}O$. What
we are discussing in the above paragraph is an open shell effect. 

To summarize, for the cases considered we get the smallest value of
$B(M1)_{spin}$ (and also $B(M1)_{physical}$) for $x=1~y=1$ and the
largest value for $x=1.5~y=0.5$ i.e. a stronger spin-orbit and a
weaker tensor interaction than what is furnished by large $G$ matrix
elements. 

{\bf (b) The Effects of the Spin-Orbit and Tensor Interactions on the
Orbital $B(M1)$}

The effect of varying $x$ and $y$ on the orbital summed strength is
more complicated than for the spin $B(M1)$. The effects are however
not insignificant. 

For $N=Z$ nuclei, we find that the effects become more important as we
go the end of the shell. For $^{20}Ne$, the effect of changing $x$ and
$y$ are very small. For the four sets of $x$ and $y$, the values of
$B_l$ are 0.95, 1.09, 0.96, and 1.10 (in the order as shown in Table
II). For $^{32}S$, the results are 2.32, 1.57, 2.17 and 1.15; and for
$^{36}Ar$ they are 1.44, 0.95, 1.33 and 0.87. We also see that
increasing the spin-orbit strength for $^{24}Mg$, $^{32}S$ and
$^{36}Ar$ decreases the summed orbital strength. Decreasing $y$ (the
tensor strength) in some cases (e.g. $^{20}Ne$) slightly increases the
orbital strength, but in most cases, it causes a decrease in this
strength. For example in $^{32}S$, going from $x=1~y=1$ to $x=1~y=0.5$
the orbital strength decreases slightly from 2.32 to 2.17; whilst from
$x=1.5~y=1$ to $x=1.5~y=0.5$ the orbital strength decreases from 1.57
to 1.15 $\mu_N^2$.

{\bf (c) The Physical $B(M1)$ -additional observations}

The dependence of the physical $B(M1)$ on $x$ and $y$ is qualitatively
similar to that of $B(M1)_{spin}$. In general, increasing $x$ and
decreasing $y$ will cause the summed strength $B(M1)_{physical}$ to
increase. 

Recalling that $\frac{g_s(isovector)}{g_l(isovector)}=9.41$, this
fact alone would cause $B(M1)_{spin}$ to be about 85 times larger than
$B(M1)_{orbital}$. However, overall in Table~II, $B(M1)_{orbital}$ and
$B(M1)_{spin}$ are comparable in magnitude. For example, for $x=1~y=1$
in $^{24}Mg$, $B(M1)_{orbital}=2.08$ and $B(M1)_{spin}=1.73$. It is
true that when we change to $x=1.5~y=0.5$ the $B(M1)_{spin}$ becomes
larger than $B(M1)_{orbital}$ (4.04 vs. 1.69), but the ratio is much
less than 85. 

These and other numbers in Table~II confirm what was said in a
previous section -that the spin matrix elements are strongly
suppressed because in deformed nuclei one moves away from the $jj$
limit towards the $LS$ $SU(4)$ limit.

{\bf (d) The Relationship of the Physical $B(M1)$ to the Sum of the
Orbital $B(M1)$ and the Spin $B(M1)$}

Let us consider the $B(M1)$ transition from the ground state $0^+$ to
any single $1^+$ state. Then the physical $B(M1)$ will be the sum of
the orbital $B(M1)$ and the spin $B(M1)$ plus their cross terms which
in principle may be positive or negative. In Table~III the spin,
orbital and physical $B(M1)$ results each represent the sum (over all
the $1^+$ states in our space) of the corresponding single-state
$B(M1)$'s. Let us now treat the $^{22}Ne$ results as one unit, summing
the $T=1~\rightarrow T=1$ (both isovector and isoscalar) and
$T=1~\rightarrow T=2$ results. Then for all the nuclei that we
consider and for all the $x,y$ combinations (all 28 cases except for
the $x=1~y=1$ case in $^{44}Ti$ with a difference of less than 1\%),
the physical $B(M1)$ is {\em less} than the sum of the spin $B(M1)$
and orbital $B(M1)$. Remarkably, in all the $^{20}Ne$, $^{22}Ne$ and
$^{44}Ti$ cases the difference is less than 3\%; for $^{24}Mg$ less
than 6\%; all these cases are in the first half of the major
shell. For the other three cases $^{12}C$, $^{32}S$ and $^{36}Ar$ (all
in the second half of major shells), the difference is still always
less than 20\%, with the physical $B(M1)$ being less than the sum of
the other two. Just to illustrate, in $^{20}Ne$ the four respective
spin $B(M1)$ plus orbital $B(M1)$ sums (i.e. for the four $x,y$
combinations) are 1.54, 2.37, 1.59 and 2.59; the corresponding
physical $B(M1)$ values are 1.53, 2.34, 1.57 and 2.54!

{\bf (e) Summed Strengths for Magnetic Dipole Excitations}

In Tables~III and IV we compare the summed strength (somewhat
arbitrarily) to the first ten states, and we also give the total
sum. We limit ourselves to the case $x=1.5,~y=0.5$. In Table~IV we
also have $x=1.5~y=0.5$ but we have $\Delta \epsilon_s=2~MeV$. 

Concerning the orbital strength we find in some cases a significant
difference between the summed strength to all states and the sum to
the first ten states. As seen in Table~III for $^{20}Ne$ the
difference  is not so large (1.10 vs. 0.90), for $^{22}Ne$
$T=1~\rightarrow T=1$ there is a factor of two difference (0.46
vs. 0.23). For $^{24}Mg$ and $^{32}S$ the differences are also large
(2.02 vs. 1.39 and 1.15 vs. 0.53 respectively). 

This is of relevance to works on relations between summed orbital
(scissors mode) strength and $B(E2)_{0_1 \rightarrow 2_1^+}$ in
deformed nuclei. What is precisely meant by the summed $M1$ orbital
strength? Experimentally it is very difficult if not impossible to
pick up the orbital ``grass'' which presumably sets in at the very
least by the time we have come to the first ten states if not
earlier. What then do the sum rules mean if experimentally not all
orbital strength is included?

{\bf (f) Comprehensive List of $M1$ Rates}

In Tables~V and VI we give a comprehensive list 
of calculated values of excitation energy ($MeV$), $B_l$, $B_\sigma$
and $B(M1)$ in units of $\mu_N^2$ for the four sets of ($x,y$) -(1,1),
(1.5,1), (1,0.5) and (1.5,0.5). In order not to drown ourselves in
details, we limit Table~V to states for which $B(M1)$ is greater or
equal to $0.3\mu_N^2$ (there is one exception in $^{20}Ne$
$x=1,~y=0.5$ where the strengths to two nearly degenerate states is
given). However, there are other states of interest which are put into
Table~VI. For example, there are states with substantial orbital
strengths and spin strengths but the destructive interference of the
orbital and spin amplitudes causes $B(M1)$ to be very small. In the
case of $^{36}Ar$, such an `invisible' state is lower in energy than
the lowest strongly excited state. 

On the whole, the behaviour in Table~V is similar to that of Table~II
where the total summed strength is given. That is to say, in going
from ($x,y$)=(1,1) to (1.5,1) the values of $B_\sigma$ and $B(M1)$
increase. In making the analysis however, we have to take note that
the fragmentation can be different for different values of ($x,y$). 

The fragmentation mentioned above is greater in some nuclei than in
others. Whereas for $^8Be$ and $^{12}C$ there is only one state with a
calculated value of $B(M1)$ greater than 0.3 $\mu_N^2$, in $^{32}S$
and in $^{36}Ar$ there are four (with $x=1.5,~y=0.5$). In $^{32}S$
states calculated to be at 10.05, 11.21, 13.01 and 13.81 $MeV$ have
values of $B(M1)$ 2.91, 5.34, 0.68 and 0.44 $\mu_N^2$ respectively. In
$^{36}Ar$ states at 11.91, 12.97, 13.73 and 17.65 $MeV$ have $B(M1)$
values of 2.62, 1.46, 0.55 and 0.49 $\mu_N^2$. For $^{36}Ar$ the
orbital strengths to these states is very small -it is mainly spin
strength. 

Table~VI shows states with very weak $B(M1)$'s but for which $B_l$
and/or $B_\sigma$ are fairly large. The destructive interference
between the orbital and spin {\em amplitudes} causes $B(M1)$ to be
very small. In the following discussion, we will be referring to
calculated results for $x=1.5,~y=0.5$ (not experiment). 

In some cases it is the lowest $1^+~T=1$ state for which this
destructive interference occurs. This is especially true for $^{36}Ar$
where the lowest calculated $J=1^+~T=1$ state for $x=1.5~y=0.5$ is at
7.80 $MeV$ with $B(M1)$ negligibly small ($<~5\times
10^{-3}\mu_N^2$). The next $1^+~T=1$ state is over 4 $MeV$ higher in
energy. This state is at 11.91 $MeV$ and has $B(M1)=2.62\mu_N^2$. The
7.80 $MeV$ state has significant $B_l$ and $B_\sigma$ values:
$0.305\mu_N^2$ and $0.320\mu_N^2$ respectively. The amplitudes are
nearly equal but of opposite sign, thus causing $B(M1)$ to nearly
vanish. 

In $^8Be$ and $^{20}Ne$, the lowest states have the largest $B(M1)$
values, but not too high above these are the `invisible' states with
substantial $B_l$ and $B_\sigma$ but negligible $B(M1)$. Thus from a
pure ($e,e'$) experiment one will miss considerable orbital strength
which is buried in these invisible states. Only with a combination of
($e,e'$) and another probe e.g. ($p,p')$ can one unravel the spin and
orbital content of the $M1$ strength. 

{\bf (g) $B(E2)$ Rates}

In Table~VII we present results for summed isoscalar and isovector
$B(E2)$ for the four different sets of ($x,y$). The general $E2$
operator is 

\[\sum_{\pi} e_pr^2Y_{2,\mu} ~+~ \sum_{\nu}e_nr^2Y_{2,\mu}\]

\noindent For the isoscalar case we take $e_p=1.5,~e_n=0.5$ and for
the isovector case $e_p=1,~e_n=-1$.

It should be noted that, for transitions $J=0^+~T=0~\rightarrow
J=2^+~T=1$, the overwhelming amount of the strength goes to the lowest
$J=2^+$ state. For example, for $^{20}Ne$ (using $x=1,~y=1$) the value
of $B(E2)\uparrow$ to the lowest $2^+$ state at 2.92 $MeV$ is 299.4
$e^2fm^4$ whilst the total sum is 315.1 $e^2fm^4$. This seems to be
true for all nuclei except $^{22}Ne$ where there is considerable
fragmentation between the first two $2^+$ states. For $x=1,~y=1$, the
value of $B(E2)$ is 198.9 $e^2fm^4$ to the $2_1^+$ state at 2.27 $MeV$
and 113.4 $e^2fm^4$ to the 3.10 $MeV$ state. The sum of these two
strengths (312.3 $e^2fm^4$) is however close to the total strength
(352.7 $e^2fm^4$). 

It should be noted that for most nuclei considered here, the
sensitivity of $B(E2)$ to $x$ and $y$ is much less than it is for isovector
$B(M1)'s$. For example, for $^{20}Ne$ the four values of
$B(E2)_{isoscalar}$ are 315.1, 313.5, 316.0 and 311.1 $e^2fm^4$
respectively. 

This can be understood from the fact that the strong central
interaction causes the low lying $2^+$ states to be dominantly $S=0$
states. The expectation values of the spin-orbit and tensor
interactions for pure $S=0$ states are zero. 

For the $^{32}S$ nucleus however, there is considerable sensitivity of
the $B(E2)_{isoscalar}$ to $x$ and $y$. For this nucleus there must be
a lot of the $S=1$ component in the ground state. 

\section{Effects of Raising the $1s$ Level with Respect to the $0d$
Level}

In our calculations for nuclei in the $1s-0d$ shell the calculated
physical $B(M1)$ values, especially with the $x=1~y=1$ interaction,
were often smaller than the corresponding measured experimental
values. This suggested to us that for these nuclei we should further
investigate one additional aspect of our calculations, namely the
energy separation of the single-particle $1s$ and $0d$ levels. From
Table~I it appears that the $1s$ level is too low in energy relative
to the centroid of the $d_{5/2}$ and $d_{3/2}$ levels. The fault may
lie with the fact that we are using harmonic oscillator wavefunctions
to evaluate matrix elements of the two-body interaction. 

To study the above effect we redid our calculations with the
$x=1.5~y=0.5$ interaction for the $1s-0d$ nuclei with the $1s$ level
raised by an {\em additional} 2 $MeV$ above the $0d$ level. This
correction is equivalent to adding a term $D\vec{l}\cdot \vec{l}$ to
the single-particle spectrum, as was done in the Nilsson model. To
calculate the best value of the energy difference to adopt, we would
have proceeded as follows. Our calculations for the single-particle 
energies in $^{17}O$ using the $x=1~y=1$ interaction (see Table~I)
yielded the following energies (in $MeV$): $\epsilon_{s_1/2}=0$,  
$\epsilon_{0d_5/2}=0.119$, $\epsilon_{0d_3/2}=5.681$. These place the
centroid of the $0d$ levels at 2.34 $MeV$ above the
$1s$. Experimentally, we find the single-particle energies (in $MeV$)
in $^{17}O$ to be $\epsilon_{d_5/2}=0$, $\epsilon_{1s_1/2}=0.871$, and
$\epsilon_{d_3/2}=5.09$. This places the $0d$ centroid at 2.03 $MeV$,
or 1.16 $MeV$ above the $1s$. Accordingly, the $1s$ should have been
raised by (2.35-1.16)=1.19 $MeV$. We chose however chose 2 $MeV$
rather than 1.19 $MeV$.

In Table~IV we present the effects of raising the $1s$ level 
relative to the $0d$ level for the $x=1.5~y=0.5$ case. In this table
we consider the summed $B(M1)$ values (physical, orbital and spin) for
transitions from the ground state to the lowest 500 $J=1^+~T=1$ states
(or a smaller number if fewer states exist in our space). From the
table, we see that the net effect of raising the $1s_{1/2}$ level is
larger in the first half of the shell. Perhaps this is because in the
second half of the shell, the $1s_{1/2}$ level plays a more minor role
in excitations, being nearly fully occupied in both the ground state
and the excited $J=1^+~T=0$ states. In the first half of the shell,
the total summed spin $M1$ strength ($B_\sigma$) increases by about
30\% in three cases: $^{20}Ne$, $^{22}Ne~ (T=1 \rightarrow T=1)$ and 
$^{24}Mg$, and by 50\% in $^{22}Ne~ (T=1 \rightarrow T=2)$. The summed
orbital isovector $M1$ strength ($B_l$) increases by 20\% and 30\% for
$^{20}Ne$ and $^{22}Ne$ respectively, but decreases by 4\% for
$^{24}Mg$. Finally the summed physical $B(M1)$'s increase by 20\% to
40\% for these nuclei ($^{20}Ne$, $^{22}Ne$ and $^{24}Mg$). This
overall increasing trend in the strength of the $B(M1)$'s is welcome
because our calculations (especially with $x=1~y=1$) typically
underestimate the experimental data there. 

However, if we look at the summed strength to the first 10 states,
there is a much bigger change in going from $x=1.5~y=0.5$ to
$x=1.5~y=0.5~\Delta \epsilon_s=2~MeV$. This can be seen from tables
III and IV. In particular, for $^{24}Mg$, the value of $B_\sigma$ to
the first 10 states (and this more or less covers the range of
experimental activity) changes from 1.79 $\mu_N^2$ to 3.38
$\mu_N^2$. This brings us close to the experimental value of 5.86
$\mu_N^2$. The increase in $B_\sigma$ when we increase the $1s$
single-particle energy can be easily understood. For a pure $(1s)^n$
ground state configuration there can be no $M1$ excitations. Thus, for
say $^{20}Ne$, when we increase $\Delta \epsilon_s$, the occupation of
$1s$ goes down and the occupation of $0d$ goes up. Thus there will be
an increase in $B_\sigma$.

One can also see this from the energy-weighted sum rule of
Kurath~\cite{kur}. He assumed that the most important part of the
Hamiltonian 
for $M1$ spin transitions is the one-body spin-orbit interaction
$-a\vec{l}\cdot \vec{s}$. Following a discussion of Zamick, Abbas and
Halemane~\cite{zah}, the isovector $M1$ operator can be written as
$\sqrt{3/16\pi}[-j_i\tau_i+(g_n-g_p+1)s_i\tau_i]$ where
$\vec{j}=\vec{l}+\vec{s}$, $g_n=-3.836~\mu_N$ and
$g_l=5.586~\mu_N$. Neglecting the $j_i\tau_i$ term, the energy
weighted sum rule of Kurath~\cite{kur} obtained by the
double-commutator method is:

\[\sum (E_n-E_0)B(M1)\uparrow=\frac{3}{32\pi}a(g_n-g_p+1)^2 
\sum_l[ln_{l+1/2}-(l+1)n_{l-1/2}]\]

\noindent where $n_j$ is the occupation of the $j$ orbit. Clearly if
$l=0$ there will be no contribution. For the $s-d$ shell, the relevant
factor is ($2n_{d5/2}-3n_{d3/2}$).  

For the $N=Z$ even-even nuclei in the second half of the shell, the
effect of raising the $1s$ level by 2 $MeV$ is less dramatic, perhaps
due to the possible reasons noted above. In $^{32}S$ and $^{36}Ar$ the
physical $B(M1)$ sums {\em decrease} by 5\% and 7\% respectively, the
orbital $B(M1)$'s increase by 4\% and 8\% and the spin $B(M1)$'s
decrease by 5\% in $^{36}Ar$ and by 9\% in $^{32}S$. 

In Table~V, and for the $x=1.5~y=0.5$ interaction, we can see the
effects of raising the $1s$ by 2 $MeV$ above the $0d$ level on the
summed $B(E2)$'s (from the $0^+$ ground state to up to 500 $2^+$
states). For the isoscalar transitions from the $J=2^+~T=0$ (which
dominate by an order of magnitude), the effect is a decrease in the
first half of the shell (by about 15\% in $^{20}Ne$, 13\% in the $T=1
\rightarrow T=1$ transitions in ${22}Ne$, and by 2\% in $^{24}Mg$) and
an increase in the second half of the shell (by 26\% for $^{32}S$ and
by 10\% for $^{36}Ar$). For the weaker isovector transitions, the
effect is smaller; it is usually a decrease (by 11\% in ${20}Ne$, 4\%
for $T=1 \rightarrow T=2$ in $^{22}Ne$, 4\% for $^{32}S$ and 7\% in
$^{36}Ar$) except for a 3\% increase in $^{24}Mg$. 

In all the five nuclei in Table~V, the isoscalar $B(E2)$ sum over all
the states is dominated by the $B(E2)$ from the ground state to the
lowest $J=2^+~T=0$ state; that state contributes between about 75\%
and 90\% of the total. By contrast, there is less dominance of a
single state in the isovector $B(E2)$ sum (from the ground state to
the $J=2^+~T=1$ states). There the state with the strongest transition
contributes about 20\% ($^{24}Mg$) to 60\% ($^{32}S$). 

\section{ Discussion of Individual Levels -Comparison
with Experiment}

We now proceed with a more systematic discussion of individual levels
and a comparison with experiment. This will involve information
contained in Tables~VIII up to XXVI.

\subsection{$^8Be$}

In $^8Be$, there is a strong $M1$ transition to the ground state from
a $J=1^+~T=1$ state at 17.6 $MeV$ from which we can infer that
$B(M1)\uparrow=0.8~\mu_N^2$. It is tempting to associate this 17.6
$MeV$ state with a scissors mode excitation. It is of interest to note
that this is the {\em only} excited state of $^8Be$ that communicates
with the ground state i.e. has an observed branch to it. The
$2^+$-to-ground $E2$ is {\em not} observed because the $2^+$ state
quickly decays into two alpha particles. We expect $^8Be$ to be a
strongly deformed nucleus. 

We find that for ($x=1,~y=1$) the strongest $B(M1)\uparrow$ to a $1^+$
$T=1$ state is to the lowest calculated state at 13.73 $MeV$ with
$B(M1)\uparrow=0.71~\mu_N^2$. The lowest $1^+~T=0$ state is at a
higher energy (14.385 $MeV$). This inverse isospin ordering is in
agreement with experiment. However the states come too low in energy
compared with the experimental values (17.6 $MeV$ and 18.5 $MeV$). 

A spin-orbit analysis shows that the (calculated) state is not purely
orbital:$B_l=0.26~\mu_N^2$ and $B_\sigma=0.15~\mu_N^2$. There is
another state calculated to be nearly 3 $MeV$ above the 13.73 $MeV$
state with $B_l=0.23~\mu_N^2$, almost as much as for the lowest
state. The value of $B_\sigma$ to this state is 0.053
$\mu_N^2$. However, destructive interference between spin and orbit
causes $B(M1)$ to be negligible ($B(M1)=0.065~\mu_N^2$). 

When we change from $x=1~y=1$ to $x=1.5~y=1$, $B_l$ changes very
little, but $B_\sigma$ increases to 0.217 $\mu_N^2$ and consequently
$B(M1)$ increases to 0.97 $\mu_N^2$. If we finally change to
$x=1.5,~y=0.5$ i.e. increase the spin-orbit and decrease the tensor,
the value of $B(M1)$ increases slightly to 1.01 $\mu_N^2$. 

It should be noted that increasing the spin-orbit interaction does not
increase the energy of the $J=1^+$ state. From $x=1~y=1$ to
$x=1.5~y=1$ the energy actually decreases from 13.73 $MeV$ to 13.06
$MeV$. This is contrary to the behaviour of a naive spherical model
where the $1^+$ state is pictured as a particle-hole state of the form
$d_{3/2}d_{5/2}^{-1}$. The energy of such a state would clearly
increase linearly with the spin-orbit strength $x$. However, since
$^8Be$ (and most of the other nuclei we will be considering here) are
strongly deformed, such `spherical' arguments do not hold water. This
problem has been discussed by Zheng and Zamick~\cite{zz}. Previous
work on $^8Be$ and $^{10}Be$ by Fayache et. al. should be noted. 

As seen from Table~VI, there is also in $^8Be$ an `invisible'
state. For $x=1.5,~y=0.5$ this state is 3.7 $MeV$ above the lowest
strong state. The calculated values of $B_l$, $B_\sigma$ and $B(M1)$
are 0.14, 0.023, and 0.050 $\mu_N^2$ respectively. There is
substantial orbital strength. Experimentally, such a state has not
been seen. 

Studies of $^8Be$ and $^{10}Be$ were previously carried out by
Fayache, Zamick and Sharma~\cite{fay1,fay2}. The study was limited to
the $x=1~y=1$ case, but the quadrupole-quadrupole interaction was also
studied. With the $Q \cdot Q$ interaction, it can be shown that the
the orbital strength $B_l\uparrow$ in $^8Be$ is
$\frac{2}{\pi}\mu_N^2=0.64\mu_N^2$. This is very close to the results 
obtained in Table~II (0.64 or 0.67 $\mu_N^2$). 

\subsection{$^{10}Be$}

In $^{10}Be$, there is no information on $M1$ excitations. There is a
strong $E2$ from the ground state to the $2_1^+$ state
$B(E2)\uparrow=52~e^2fm^4$. Raman $et. al.$ deduce from this a
deformation parameter $\beta=1.13$~\cite{raman}. The experimental
result is somewhat smaller than the isoscalar result of Table~VII (71
$e^2fm^4$). It should be remembered that the theoretical results were
obtained with effective charges $e_p=1.5~e_n=0.5$. There is some
evidence that in the $0p$ shell, the effective charges should be
somewhat smaller. 

As shown by Fayache $et. al.$~\cite{fay2}, with a $Q\cdot Q$
interaction one could obtain analytic results for the isovector
orbital strength $B_l$ in $^{10}Be$. The results were: 

\[B_l(0^+~T=1 \rightarrow 1^+~T=1)=\frac{9}{32\pi}\mu_N^2= 0.0895
\mu_N^2\]

\noindent and

\[B_l(0^+~T=1 \rightarrow 1^+~T=2)=\frac{15}{32\pi}\mu_N^2= 0.1492 
\mu_N^2\]

\noindent The total orbital strength ($T=1+T=2$) of
$\frac{24}{32\pi}\mu_N^2$ is only 3/8 of the total orbital strength in
$^8Be$. 

As seen in Table~II, the results for isovector $B_l$ for all four
cases of the $(x,y)$ interaction interaction seem to be in close
accord with the analytic results of the $Q \cdot Q$ interaction
above. 

From Table~V we note that the two lowest lying states fro $x=1~y=1$
are calculated to be at 6.14 $MeV$ and 7.68 $MeV$ with values of
$B(M1)$ of $1.1~10^{-2}~\mu_N^2$ and $1.85~\mu_N^2$
respectively. These states have negligible orbital content -they are
spin modes not orbital modes. The calculated $T=1$ and $T=2$ isovector
orbital strengths (i.e. scissors modes)~\cite{fay2} are at much higher
energies $\approx 19 ~ MeV$. With a $Q \cdot Q$ interaction, the $T=1$
and $T=2$ orbital strengths are degenerate. For $x=1.5~y=0.5$ we get
two states with significant $B(M1)$ strength. The energies are 8.45
$MeV$ and 9.26 $MeV$, and the values of $B(M1)\uparrow$ are 0.77
$\mu_N^2$ and 2.70 $\mu_N^2$. The orbital content is still
negligible. 

In Table~VI we list for $^{10}Be$ those states which contain most of
the orbital strength. There are both $T=1$ and $T=2$ states, and for
$x=1.5~y=0.5$ they range in energy from about 16 $MeV$ to 24
$MeV$. The orbital strength is at a much higher energy than the spin
strength for $^{10}Be$.

\subsection{$^{12}C$}

As shown in Table~VIII, there are two well-studied states in $^{12}C$:
the $1^+~T=0$ state at 12.71 $MeV$ and the $1^+~T=1$ (isovector) state
at 15.11 $MeV$. As mentioned in the introduction, the ratio 
$(|\frac{g_s(vector)}{g_s(scalar)})^2=81$ so we expect the $1^+~T=1$
state to be excited much more strongly than $T=0$ and indeed in the
calculation this is the case. For $x=1~y=1$, we get 

\[\frac{B(M1)\uparrow(g.s. \rightarrow
1^+~T=1)}{B(M1)\uparrow(g.s. \rightarrow
1^+~T=0)}=\frac{0.856}{0.0026}=330\]

The lowest calculated $1^+,~T=1$ state in $^{12}C$ is dominantly a
spin state. For $x=1~y=1$ the values of $B(M1)$ (in $\mu_N^2$) are
$B_l=0.0314$, $B_\sigma=0.584$ and $B(M1)=0.886$. For the other
extreme ($x=1.5~y=0.5$), we obtain 
$B_l=6.91~10^{-5}$, $B_{\sigma}=2.571$ and $B(M1)=2.54$. This analysis
confirms the assumption made in testing the CVC theory by A. Richter
at. al. -that the the 15.1 $MeV$ state is indeed an almost pure spin
state. Only then can a relation between the isovector $M1$ and a
corresponding Gamow-Teller matrix element from beta decay be made. 

Where there are orbital (scissors mode) excitations, the analysis is
tricky. For $x=1.5~y=0.5$, the next excited $J=1^+~T=1$ state is at
17.46 $MeV$, and it has $B_l=0.28~\mu_N^2$ and
$B_{\sigma}=0.074~\mu_N^2$. But there is a large cancellation between
spin and orbit so that $B(M1)$ is only 0.068 $\mu_N^2$. One would need in
addition to electromagnetic excitation another probe e.g. proton
scattering to unearth the substantial orbital strength at 20.19 $MeV$
($B_l=0.17~\mu_N^2$, $B_{\sigma}=0.0052~\mu_N^2$,
$B(M1)=0.113~\mu_N^2$). 

We now consider the ($x,y$) dependence of $B(M1)$. For the four values
(1,1), (1.5,1), (1,0.5) and (1.5,0.5), the values of $B(M1)$ (in
$\mu_N^2$) are 0.89, 1.89, 1.12 and 2.54 respectively. The
experimental value is 2.63(8) $\mu_N^2$ so we see that we need the
full machinery (larger spin-orbit and weaker tensor interactions) to
come even close to the large measured value. 
Again, as in $^8Be$, the energies of the $1^+$ states are somewhat too
low, and changing $x$ and $y$ does not solve the problem. Thus for
$x=1.5~y=0.5$ the $1^+$ states are calculated to be at 11.8 $MeV$
($T=0$) and 13.1 $MeV$ ($T=1$). 

A comparison of experiment and theory for $B(E2)\uparrow$ in $^{12}C$
is made in Table~IX. By far the largest $B(E2)$ is to the first $2^+$
state at 4.44 $MeV$. With the `standard' effective charges
$e_p=1.5~e_n=0.5$, the calculated values of $B(E2)$ range from 70 to
80 $e^2fm^4$ -they are not very sensitive to variations in $x$ and
$y$. The experimental value is 40 $e^2fm^4$. It appears from the
results here and in $^8Be$ that the effective charges in the $0p$
shell should be somewhat smaller than the standard value. For
$^{12}C$, and for all even-even nuclei considered here, the $B(E2)$
$T=0 \rightarrow T=0$ is proportional to $(e_p+e_n)^2$ and the $B(E2)$
$T=0 \rightarrow T=1$ to $(e_p-e_n)^2$. 

\subsection{$^{20}Ne$}

The experimental data for $^{20}Ne$ is contained in Tables~X and XI. 
In $^{20}Ne$ there is a strong $M1$ excitation to a $J=1^+~T=1$ state
at 11.26 $MeV$. The experimental values to this state (in $\mu_N^2$) are
$B_l=0.52(15)$, $B_{\sigma}=0.49(6)$ and $B(M1)=2.02(36)$. The
calculated values of $B(M1)$ for $x=1~y=1$ are $B_l=0.43$,
$B_{\sigma}=0.06$ and $B(M1)=0.83$. For $x=1.5~y=0.5$, we get
$B_l=0.47$, $B_{\sigma}=0.35$ and $B(M1)=1.63$. We thus get
substantially better values of $B_{\sigma}$ and hence of $B(M1)$ when
we increase the spin-orbit interaction and decrease the tensor. 

But, as mentioned previously, we appear to get the `correct'
spin-orbit splitting in $^{17}O$ with a value of the spin-orbit
strength $x=1$. This gives a $d_{3/2}-d_{5/2}$ single-particle
splitting of 5.2 $MeV$. For $x=1.5$ the value would be $5.2 \times
1.5=7.8~MeV$. Thus the $M1$'s in $^{20}Ne$ (and in other $s-d$ shell
nuclei as well) act as if the spin-orbit splitting is much larger than
what is obtained from energy splittings for $A=17$. 

When we further move the $1s$ level up by 2 $MeV$ ($\Delta
\epsilon_s=2 MeV$) relative to $0d$, the values of $B_l$, $B_\sigma$
and $B(M1)$ to the strongest state (calculated to be at 9.76 $MeV$)
become (in units of $\mu_N^2$) 0.51, 0.62 and 2.26 respectively. We
see that $B_\sigma$ and hence $B(M1)$ increase and indeed, for
$B(M1)$, we are within the errors bars of the experiment. 

The dominant $E2$ strength goes to the $2_1^+$ state at 1.63
$MeV$. The measured $B(E2)\uparrow$ is 340(22) $e^2fm^4$ whilst the
calculated values (with $e_p=1.5,~e_n=0.5$) are about 300 $e^2fm^4$
-in good agreement. As in $^{12}C$, there is very little sensitivity
of $B(E2)$ to $x$ and $y$. This could be understood if the extreme
$LS$ coupling limit was valid. If the ground state is $L=0~S=0$ and
the $2^+$ state $L=2~S=0$ then the tensor and spin-orbit interactions
would have no effect -the expectation values will vanish for $S=0$
states. With $\Delta \epsilon_s=2 MeV$, the $B(E2)$ to the $2_1^+$
state decreases to 265 $e^2 fm^4$. There is some loss of collectivity
when the $1s$ level is raised in energy. 

\subsection{$^{22}Ne$}

Experimentally (see Table~XII), the $B(M1)$'s in $^{22}Ne$ are as
follows: 0.43(17) $\mu_N^2$ to a state at 5.33 $MeV$, 1.36(56)
$\mu_N^2$ to a state at 6.85 $MeV$ and 1.80(7) to a state at 9.18
$MeV$. The summed experimental strength to the three states is 3.59
$\mu_N^2$. For the four ($x,y$) values we get the sum to be 1.62
$\mu_N^2$, 2.76 $\mu_N^2$, 1.96 $\mu_N^2$ and 3.38 $\mu_N^2$. We
clearly get better results when $x=1.5$ (second and fourth cases), and
we get the best results with $x=1.5,~y=0.5$. In a later section, we
shall see that even better results are obtained by increasing the $1s$
level ($\Delta \epsilon_s=2MeV$). 

In the absence of a spin-orbit interaction, the nucleus $^{22}Ne$ is
triaxial in a Hartree-Fock calculation e.g. with a Skyrme
interaction. Introducing a spin-orbit interaction tends to make this
and all other nuclei axially symmetric. Depending on what parameters
are used, the calculation can go either way -triaxial or axially
symmetric. 

We can determine the degree of triaxiality by looking at
$B(E2)\uparrow$ from the ground state. In the axial case, the lowest
$2^+$ state will be a member of a $K=0$ ground state band and the next
one of a different band with $K=2$. Thus $B(E2)~0_1 \rightarrow 2_1$
will be much stronger than $B(E2)~0_1 \rightarrow 2_2$. In the
triaxial case, there will be much $K$ mixing and we can get strong
$B(E2)$'s to both the $2_1$ and $2_2$ states. 

Experimentally, as shown in Table~XIII, the $B(E2)$ to the first $2^+$
state is {\em much} larger than to the second. The respective values
are 300 $e^2fm^4$ and 12 $e^2fm^4$.  These results favor axial
symmetry for $^{22}Ne$. 

In the shell model calculations with the ($x,y$) interaction, we also
see a strong dependence of the ratio $B(E2)_2$ to $B(E2)_1$ with the
strength of the spin-orbit interaction, and indeed the results are
consistent with those of Hartree-Fock calculations. 

For $x=1$ (i.e. the free-space spin-orbit interaction) and $y=1$, we
get strong $B(E2)$'s to both the first and second $2^+$ states. The
respective values are 199 $e^2fm^4$ and 113 $e^2fm^4$. When we
increase $x$ to 1.5 (keeping $y=1$) the second $B(E2)$ shrinks
considerably. The values are now $B(E2)_1$=289  $e^2fm^4$ and
$B(E2)_2$= 7.5  $e^2fm^4$. 

Thus the $B(E2)$ data, just like the $M1$ data, favour a larger
spin-orbit interaction than in free space. Thus we have the
interesting observation that the absence of triaxiality throughout the
periodic table may be due to the fact that the spin-orbit interaction
in a nucleus gets enhanced by medium modifications. As mentioned in
the introduction, there are several models for such a modification -a
Dirac effective mass in the Walecka model, and three-body interactions
due to meson degrees of freedom in the Argonne-Illinois model.

\subsection{$^{24}Mg$}

In $^{24}Mg$ we have the ($e,e'$) results of Richter
et. al.~\cite{rw} as shown in Table~XIV. They obtain a summed
$B(M1)$ of 4.85(36) $\mu_N^2$ up to a maximum excitation energy
$E_{max}=11.4~MeV$, and the sum increases to 5.84(40) $\mu_N^2$ for
$E_{max}=15~MeV$. 

In our calculation the summed $B(M1)$ strength to {\em all} states in
the $sd$ shell for the four values of ($x,y$) are 3.78, 4.95, 3.87 and
5.43 $\mu_N^2$. The sums for the first 10 states (which have maximum
excitation energies which vary from 15 to 16 $MeV$ i.e. in the same
ballpark as the experiment) are smaller: 2.80, 3.28, 2.90 and 3.45
$\mu_N^2$. We thus find that the experimental sum is larger than the
theoretical sum by a factor
$\frac{exp}{free}=\frac{5.43}{3.45}=1.57$. The USD interaction of 
Wildenthal also gives a ratio bigger than one but it is much smaller
than ours: 1.11(8). 

However, with $\Delta \epsilon_s=2 MeV$ in the $x=1.5~y=0.5$ case, the
situation improves a great deal. The total sum increases to 6.44
$\mu_N^2$, and the sum to the first ten states increases to 4.58
$\mu_N^2$. 

Experimentally, two low-lying states absorb most of the $M1$
strength. We have $B(M1)\uparrow=1.5\mu_N^2$ to a 9.96 $MeV$ state and
2.5 $\mu_N^2$ to a 10.71 $MeV$ state. In the calculation however, for
say ($x,y$)=(1.5,0.5), the values of $B(M1)$ greater than 0.3
$\mu_N^2$ are 2.11 $\mu_N^2$ to a 9.32 $MeV$ state and 0.64 $\mu_N^2$
to a 9.47 $MeV$ state. Since in the shell model calculation these two
states are nearly degenerate, it is hard to get the strengths just
right. The experimental summed strength to the two states is 4
$\mu_N^2$ whilst the calculation gives 2.75 $\mu_N^2$.

In $^{24}Mg$ we also have results for $B_\sigma$ from proton inelastic
scattering by Crawley {\em et. al.}~\cite{craw} in ($p,p'$) reactions as
shown in Table~XV (see also Table~XVI). The summed 
experimental strength for states ranging in energy from 8.865 to
14.267 $MeV$ is 5.86 $\mu_N^2$. The first two $1^+$ states at 8.865
and 9.820 are weakly excited: $B_\sigma=0.042\pm 0.011$ and $0.260 \pm
0.040$ $\mu_N^2$ respectively. The next two states at 9.962 and 10.711
$MeV$ are strongly excited: $B_\sigma=1.160\pm 0.180$ and $3.180 \pm
0.300$ $\mu_N^2$ respectively. The summed experimental strength for
the above-mentioned energy range is 5.862 $\mu_N^2$.

Theoretically, the entire summed strength of $B_\sigma$ for the four
values of ($x,y$) are respectively 1.73, 3.35, 1.94 and 4.04
$\mu_N^2$. For the first ten states we get (for $x=1.5,~y=0.5$) $\sum
B_\sigma= 1.79 \mu_N^2$. This is considerably smaller than the
experimental sum. But with $\Delta \epsilon_s=2 MeV$ the total
$B_\sigma$ sum is 5.45 $\mu_N^2$ and to the first ten states 3.38
$\mu_N^2$. 

As to individual levels and for $x=1.5~y=0.5$, we get nearly
degenerate states at 9.325 and 9.466 $MeV$ with $B_\sigma=1.02$ and
0.27 $\mu_N^2$ i.e. a summed strength of 1.29 $\mu_N^2$. The summed
experimental strength $\sum B_\sigma$ for the two strong states is
4.34 $\mu_N^2$. With $\Delta \epsilon_s=2 MeV$ the corresponding
summed $B_\sigma$ strength is 2.41 $\mu_N^2$, somewhat closer to
experiment. 

\subsection{$^{28}Si$}

From the work of L\"{u}ttge {\em et. al.}~\cite{lut} as shown in Table
XVIII, the summed $B(M1)\uparrow$ strength in $^{28}Si$ up to an
excitation energy of 18 $MeV$ is 7.53 $\mu_N^2$. There is considerable
fragmentation with the three strongest states at energies of 10.901,
11.445 and 12.331 $MeV$ having $B(M1)$'s of $0.90 \pm 0.02 \mu_N^2$,
$4.42 \pm 0.20 \mu_N^2$ and $0.87 \pm 0.06 \mu_N^2$ respectively. From
the work of Crawley {\em et. al.} (Table~XIX), the summed $B_\sigma$
strength up to an excitation energy of 15.8 $MeV$ is 9.51
$\mu_N^2$. The $B_\sigma$ strength to the strongest state at 11.45
$MeV$ is 3.32 $\pm$ 0.24 $\mu_N^2$ as determined by Crawley {\em
et. al.}~\cite{craw}. 

For $x=1~y=1$, our summed $B(M1)_{isovector}$ is 4.44 $\mu_N^2$ with
$\sum B_l=2.24 \mu_N^2$ and $\sum B_\sigma=2.39 \mu_N^2$. Our result
is significantly less than experiment. The experimental $B(M1)$ sum to
the first ten states is 7.35 $\mu_N^2$ whereas our sum is only 2.93
$\mu_N^2$. The calculated $B(M1)$ strength to individual states for
$x=1 ~y=1$ are 1.35, 0.49, 0.39 and 0.35 $\mu_N^2$ at 10.31, 11.78,
12.97 and 14.54 $MeV$ respectively. We seem to be missing the very
strong state with an experimental $B(M1)$ of 4.42 $\mu_N^2$. 

The results are dramatically improved when we go to $x=1.5$. The value
of $B(M1)$ to the single strongest state for the cases ($x,y$)=(1,1)
and (1.5,0.5) are respectively 1.35 $\mu_N^2$ and 3.90
$\mu_N^2$. There is almost a factor of three increase and we come
close to the experimental value of 4.42 $\mu_N^2$. The corresponding
change in $B_\sigma$ shown in Tables~XIX, XX and XXI is from 0.86
$\mu_N^2$ to 2.26 $\mu_N^2$, in much better agreement but still lower
than the experimental value. From ($x,y$)=(1,1) to (1.5,0.5) the {\em
total summed strength} (Table~II) for $B(M1)$ changes from 4.44 to
6.38 $\mu_N^2$, whilst the total summed $B_\sigma$ strength changes
from 2.39 to 6.79 $\mu_N^2$. If we limit the sum to the first ten
states (Table~III), the change in $B_\sigma$ is from 1.29 to 4.76
$\mu_N^2$. In the latter case, $B_\sigma$ increases by a factor of
3.7. 

We nest consider the case ($x,y$)=(1.5,0.5) with $\Delta \epsilon_s =
2 MeV$. The values of $B(M1)$ and $B_\sigma$ to the single strongest
state (calculated to be at 10.03 $MeV$) are 4.61 $\mu_N^2$ and 4.26
$\mu_N^2$ respectively. The total summed strength for $B(M1)$ and
$B_\sigma$ are now increased to 8.94 and 9.04 $\mu_N^2$, whilst to the
first ten states the sums are 7.08 and 7.24 $\mu_N^2$. 

In contrast to $B_\sigma$, there is a substantial decrease in $B_l$
when we go from ($x,y$)=(1,1) to (1.5,0.5). The respective values for
the total $B_l$ strengths are 2.24 $\mu_N^2$ and 1.46 $\mu_N^2$
respectively. With ($x,y$)=(1.5,0.5) and $\Delta \epsilon_s = 2 MeV$,
there is a further decrease to 1.36 $\mu_N^2$. 

\subsection{$^{32}S$}

The results are shown in Table~XXII. The most clearcut data in
$^{32}S$ appears for $B_\sigma$ by Crawley {\em
et. al.}~\cite{craw}. The summed strength is 10.75 $\mu_N^2$ and the
strongest 
excited states and the corresponding ($B_\sigma$ in $\mu_N^2$) are:
8.13 $MeV$ (1.46 $\pm$ 0.19), 11.13 $MeV$ (4.08 $\pm$ 0.53) and 11.63
$MeV$ (2.38 $\pm$ 0.35). For $x=1.5~y=0.5$ our summed $B_\sigma$
strength from table XXIV is 11.16 $\mu_N^2$ to the first ten states
and 12.03 $\mu_N^2$ to all states. The agreement with experiment is
reasonable. Our strongest state, however, has a $B_\sigma$ strength of
5.98 $\mu_N^2$ at 11.21 $MeV$. 

When for the case $x=1.5~y=0.5$ case we also use $\Delta \epsilon_s = 2
MeV$, the value of $B_\sigma$ to the first ten states drops somewhat
to 9.45 $\mu_N^2$, and the total sum also drops to 10.91
$\mu_N^2$. The strongest state calculated to be at 10.25 $MeV$ now has
$B_\sigma=4.20 \mu_N^2$, a significant drop from 5.98 $\mu_N^2$. 

Note that this is the first time we get a drop in $B_\sigma$ when
$\Delta \epsilon_s = 2 MeV$ is used; for $^{20}Ne$, $^{22}Ne$,
$^{24}Mg$ and $^{28}Si$ there was a rise. Note further that our
results with this drop are in better agreement with experiment. 

The $B(M1)$ data is a bit complicated. In Table~XXII from the results
reported in the paper of Crawley {\em et. al.}~\cite{craw} only a few
states are seen, with a summed strength of 5.43 $\mu_N^2$. However,
there is other data by Petraitis {\em et. al.}~\cite{petr} which shows
strength to many other states. The summed Peraitis strength is 4.21
$\mu_N^2$. However, he does not have the state at 8.13 $MeV$ 
($B(M1)\uparrow=1.14 \pm 0.18~\mu_N^2$), and disagrees on the $B(M1)$
to the 11.13 $MeV$ state for which Petraitis gets 1.24 $\pm$ 0.13
$\mu_N^2$ but ref.~\cite{craw} reports 2.40 $\pm$ 0.22 $\mu_N^2$.

We feel the best estimate at the moment is to take all the $M1$ states
of ref.~\cite{craw} and add to that the strength of Petraitis
{\em et. al.} to states not found in ref.~\cite{craw}. Thus we
get schematically strength reported in ref.~\cite{craw} plus
(strength reported by Petraitis {\em et. al.} minus strength to states
at 9.66 $MeV$, 11.16 $MeV$ and 11.85 $MeV$). When all this is done, we 
get a summed $B(M1)$ strength of 7.61 $\mu_N^2$. Our calculated
$B(M1)$ strength (with $x=1.5~y=0.5$) is 10.98 $\mu_N^2$ to all
states. With $\Delta \epsilon_s = 2 MeV$, this {\em drops} to 10.12
$\mu_N^2$, whilst the summed strength to the first ten states (i.e. up
to 15.1 $MeV$ excitation energy) is 8.81 $\mu_N^2$. 

It is difficult to draw a definitive conclusion for $^{32}S$ because
two sets of $M1$ data are not in accord. From the $B_\sigma$ data, it
appears that we need an enlarged spin-orbit interaction. For $x=1~y=1$
the sum $B_\sigma$ is only 2.73 $\mu_N^2$ whereas for $x=1.5~y=0.5$ it
is 12.03 $\mu_N^2$, closer to the experimental sum of 10.75
$\mu_N^2$. However, the $M1$ data supports a somewhat weaker
spin-orbit and/or a stronger tensor interaction. Again, from Table~II, 
the case $x=1.5~y=1$ gives results of 8.69 $\mu_N^2$ for $B(M1)$ and
8.52 $\mu_N^2$ for $B_\sigma$. 

It is significant that raising the $1s$ level relative to $0d$ causes
a rise in $B_\sigma$ for $^{20}Ne$, $^{24}Mg$ and $^{28}Si$, whilst
for $^{32}S$ and as we shall see $^{36}Ar$ it causes a decrease in
$B_\sigma$. 

\subsection{$^{36}Ar$}

For $^{36}Ar$, we refer to Tables~XXVII, XXVIII and XXIX. 
Because of the large error bars on the ($e,e'$) experiment, we will
concern ourselves here with the summed strength only. The measured
value is 2.65 $\pm$ 0.1 $\mu_N^2$. 

The total summed strengths for the four values of ($x,y$) are 3.35,
5.19, 3.72, 5.71 $\mu_N^2$ respectively. The sums to the first
ten states are 2.98, 4.82, 3.58 and 5.50 $\mu_N^2$ respectively. They
are almost the same as the complete sums. 

The Foltz work~\cite{foltz} contains a theoretical discussion of shell
model calculations using the Wildenthal interaction~\cite{bw2}. Even
with this phenomenological interaction there is a difficulty in the
sense that the theory predicts a strong excitation of
$B(M1)=1.3~\mu_N^2$ at 10.3 $MeV$, whereas experimentally there is
very strong fragmentation. Also the theory with bare operators
predicts too large a summed strength (5.13 $\mu_N^2$) for all the
strength, and 4.73 $\mu_N^2$ for the strength below 15 $MeV$. However,
using effective operators, whose values are consistent with the
fundamental calculations of Towner and Khanna~\cite{tk}, the summed
strength is reduced and is closer to experiment. The above two numbers
now become 3.54 $\mu_N^2$ to all states and 2.86 $\mu_N^2$ to states
below 15 $MeV$. 

Thus the results we obtain with $x=1.5~y=0.5$, although still too
large, are not so different from those obtained in the theoretical
discussion of Foltz et. al.\cite{foltz}. As seen in Tables~III and IV,
we get summed strengths of 5.71 $\mu_N^2$ to all states and 5.50
$\mu_N^2$ to the first ten states. With $\Delta \epsilon_s = 2 MeV$,
the corresponding numbers are 5.48 and 5.10 $\mu_N^2$. 

\subsection{$^{44}Ti$}

There are no $M1$ data for $^{44}Ti$ so we concentrate on $B(E2)$'s
-these are shown in Tables~XXX and XXXI. The experimental value of
$B(E2)\uparrow$ is 540(140) $e^2fm^4$ but the calculated values (with
$e_p=1.5~e_n=0.5$) are much larger -from 908 to 925 $e^2fm^4$ for
various combinations of $x$ and $y$.

The probable reason for the calculated over-collectivity is that the
$1p$ level is too low in energy relative to $0f$. This is confirmed in
calculations in which we raise the $1p$ centroid by 2 $MeV$ relative
to the $0f$ centroid. In that case, the value of $B(E2)\uparrow$ is 
reduced to 710 $e^2fm^4$ (for $x=1.5~y=0.5$).

\section{Closing Remarks}

In this work we simulate a realistic $G-$matrix by the interaction
$V_c(r)+xV_{s.o}+yV_t$ where $x=1,~y=1$ gives us matrix elements close
to those obtained in a non-relativistic approach. We must emphasize at
first that we use the same interaction to calculate single-particle
energies as is used for the valence interactions. In this, we differ
from most other calculations in which a hybrid approach is taken -most
often the single-particle energies are taken from experiment. We feel
however that one is not really testing the interaction. Later we relax
this condition somewhat by shifting the $1s$ single-particle energy
relative to $0d$. 

In an open-shell nucleus the tensor interaction has the opposite
effect of the spin-orbit interaction~\cite{wong}. That is to say when 
the strength of the tensor interaction is {\em decreased}, the summed
spin strength $\sum B_\sigma$ {\em increases}. There seems to be a
phenomenological need for a weaker tensor interaction in a nucleus
than in free space, and several arguments have been proposed to
justify this. At the simplest level, there is the `self-weakening
mechanism'~\cite{zzf,zzfm}. This simply means that the effects of
higher-shell admixtures (i.e. $\Delta N=2$ excitations and beyond) can
be simulated in the valence space by using a weaker tensor
interaction. This is a purely nuclear structure justification, and the
calculations for these admixtures can be carried out either
perturbatively or in large shell model spaces. There is also the
`universal scaling' idea of Brown and Rho~\cite{brown} that all
mesons, except for the pion, are less massive inside a nuclear medium
than in free space. This leads to a larger range for the contribution
to the tensor interaction of $\rho$-exchange. This then causes a
greater cancellation between the $\rho$- and $\pi$-exchange
contributions to the tensor interaction, leading to the overall
weakening of the tensor interaction. 

In our calculations, $B_\sigma$ does indeed increase when the tensor
interaction is cut in half. Although the effect is not so large as
increasing the spin-orbit interaction (from $x=1$ to $x=1.5$), overall
the effects seem beneficial with the possible exception of $^{36}Ar$.

It should be further noted that the tensor interaction seems to have
very little effect on the $B(E2)~0_1 \rightarrow 2_1$. This is
undoubtedly because for these states one is approaching the $LS$
limit, and the wavefunctions are dominantly $S=0$. The tensor
interaction vanishes for $S=0$ states. To our mind, this means that
one can control $B_\sigma$ to some extent without affecting the
low-lying spectroscopy. Note that for the $N=Z$ nuclei considered
here, the isovector transition $B_\sigma$ ($J=0~T=0~\rightarrow J=1
T=1$) is proportional to the Gamow-Teller transition $B(GT)$. It has
often been said that $B(GT)$ has to be quenched and there are even
more specific statements that the $GT$ coupling constant which is
1.251 in free space should be equal to unity inside the nucleus. Our
calculations here cast some doubt on such detailed conclusions. First
of all, the famous sum rule $B(GT)_+=3(N-Z)+B(GT)_-$ puts no
restrictions whatsoever on $B(GT)$ for an $N=Z$ nucleus- one only gets
the result $B(GT)_+=B(GT)_-$ which is simply isospin
conservation. Secondly, we know that $B_\sigma$ isovector and $B(GT)$
will vanish in the $SU(4)$ limit for $N=Z$ nuclei. At the other limit
of $jj$ coupling, $B_\sigma$ and $B(GT)$ are much too large. So it
seems we can get almost any answer we want. Indeed, for $x=1~y=1$ we
get too small a $B(GT)$ and if we really believed in this interaction
we could conclude that $B(GT)$ had to be {\em enhanced}. On the other
hand, if we had chosen $x=2~y=1$ (a very large spin-orbit
interaction), we could conclude that $B(GT)$ had to be quenched. 
Perhaps a case can be made that the 3($N-Z$) part of the sum rule is
quenched but not the  $B(GT)$ part. 

On the other hand, in the work of Richter, Weiss, Hausser and
Brown~\cite{rw}, the discussion of quenching pertains to the ratio of 
$B(M1)_\sigma$ to $B(GT)$, and here one is on much firmer ground. By
taking the ratio the uncertainties of the details of the nuclear
interaction largely disappear, and the conclusion that $B(GT)$ is more
quenched than $B(M1)$ is certainly valid. 

In the $0p$ shell, where there is only one $l$ orbit, it is easier to
draw conclusions. With $x=1~y=1$, we get a $B(M1)$ of only 0.89
$\mu_N^2$ to the 15.11 $MeV$ state in $^{12}C$-the experimental value
is 2.63(8) $\mu_N^2$. For $x=1.5~y=1$, the value is 1.89 $\mu_N^2$,
but only when we both increase the spin-orbit and decrease the tensor
interaction ($x=1.5~y=0.5$) do we get a $B(M1)$ comparable with
experiment (2.55 $\mu_N^2$). 

In the $s-d$ shell, things are somewhat more complicated by the fact
that there are two $l$ orbits. Our initial calculation (in which the 
$0d_{5/2}$, $1s_{1/2}$ and $0d_{3/2}$ single-particle energies are
calculated with the same ($x,y$) interaction as is used in the valence
space) puts the centroid of the $1s$ level too low in energy relative
to that of $0d$. This has the effect of making $B_\sigma$ too small in
the lower half of the $s-d$ shell e.g. $^{20}Ne$, $^{22}Ne$, $^{24}Mg$
and  $^{28}Si$. By moving the $1s$ level up by 2 $MeV$ we increase
$B_\sigma$ significantly for these nuclei, and as a bonus, in the
upper half of the $s-d$ shell i.e. $^{32}S$ and $^{36}Ar$ $B_\sigma$
gets slightly decreased, and this appears all to the good. 

We find that for most nuclei, the non-relativistic $G-$matrix, as
given by the parameters $x=1~y=1$ gives much too small values for the
spin excitation rates ($B_\sigma$). By increasing the two-body
spin-orbit term from $x=1$ to $x=1.5$ we can greatly improve the
situation. Microscopic justification for this came from the Dirac
phenomenology in which a Dirac effective mass $m^*$ is introduced, and
is said to arise from the exchange of $\sigma$ mesons between
nucleons. The spin-orbit interaction gets increased by a factor
$\frac{m}{m^*}$ (however $m^*$ is a function of $r$ and so the effect
should be greater in the interior than on the surface of the
nucleus). There is an alternate three-body approach~\cite{wal} which
also gives an enhanced spin-orbit splitting. 

Lastly, we have here introduced a new topic -the cross-correlation
between the spin and orbital parts of $B(M1)$. In our calculations, we
find that the the total sum $\sum B(M1)$ is very close to $\sum
B_\sigma$ + $\sum B_l$, indicating that the sum of the cross terms
nearly vanishes. 

\section{Acknowledgements}

Y. Sharon received support on his sabbatical from Richard Stockton 
College of New Jersey. This work was supported by a D.O.E. grant
DE-FG05-86ER-40299 and by the German Federal Ministry of Education,
Research and Technology (BMBF) under contract No. 06DA6651. One of us
(L.Z.) acknowledges support from the Alexander Von Humboldt
Foundation. We thank Alex Brown for useful communications.

\pagebreak

\pagebreak
\narrowtext
\begin{small}
\begin{table}
\caption{Single-Particle Energies Calculated with the ($x,y$)
interaction ($x=1,~y=1$) in a small space ($0\hbar\omega$) and a large
space $(0+2)\hbar\omega$.}
\begin{tabular}{llll}
Nucleus & State & $0\hbar\omega$ & $(0+2)\hbar\omega$\\
\tableline
$^5He$  & $3/2^-$ & 0 & 0\\
        & $1/2^-$ & 3.375 & 2.224\\
\tableline
$^{15}O$ & $1/2^-$ & 0 & 0\\
         & $3/2^-$ & 5.063 & 5.679\\
\tableline
$^{17}O$ & $1/2^+$ & 0 & 0\\
         & $5/2^+$ & 0.119 & 1.430\\
         & $3/2^+$ & 5.681 & 7.092\\
\tableline
$^{39}Ca$ & $3/2^+$ & 0 & \\
          & $1/2^+$ & 5.457 &\\
          & $5/2^+$ & 6.973 &\\
\tableline
$^{41}Ca$ & $3/2^-$ & 0 & \\
          & $7/2^-$ & 0.940 &\\
          & $1/2^-$ & 2.458 &\\
          & $5/2^-$ & 7.966 &\\
\end{tabular}
\end{table}

\begin{table}
\caption{The Summed isovector $M1$ strength in $\mu_N^2$ for various strengths
($x,y$) of the spin-orbit and tensor interaction}
\begin{tabular}{llllll}
Nucleus & $x$ & $y$ & $B(M1)$ & $B_l$ & $B_\sigma$\\
\tableline
$^8Be$   & 1   & 1   & 1.05 & 0.67 & 0.38\\
         & 1.5 & 1   & 1.28 & 0.64 & 0.65\\
         & 1   & 0.5 & 1.06 & 0.67 & 0.41\\
         & 1.5 & 0.5 & 1.33 & 0.64 & 0.73\\
\tableline
$^{10}Be$ & 1  & 1   & 2.09 & 0.111(0.025) & 1.62\\
$T=1~\rightarrow T=1$ & 1.5 & 1 & 3.17 & 0.105(0.036) & 2.64\\
                      & 1   & 0.5 & 2.46 & 0.111(0.028) & 2.03\\
                      & 1.5 & 0.5 & 3.55 & 0.103(0.043) & 3.11\\\\
$^{10}Be$ & 1  & 1   & 0.060  & 0.151 & 0.093\\
$T=1~\rightarrow T=2$ & 1.5 & 1 & 0.029 & 0.144 & 0.189\\
                      & 1   & 0.5 & 0.046 & 0.152 & 0.104\\
                      & 1.5 & 0.5 & 0.050 & 0.142 & 0.246\\
\tableline
$^{12}C$ & 1 & 1 & 1.42 & 0.60 & 0.91\\
         & 1.5 & 1 & 2.29 & 0.50 & 2.07 \\
         & 1 & 0.5 & 1.58 & 0.59 & 1.13 \\
         & 1.5 & 0.5 & 2.85 & 0.46 & 2.82\\
\tableline
$^{20}Ne$ & 1 & 1 & 1.53 & 0.95 & 0.59\\
          & 1.5 & 1 & 2.34 & 1.09 & 1.28\\
          & 1 & 0.5 & 1.57 & 0.96 & 0.63\\
          & 1.5 & 0.5 & 2.54 & 1.10 & 1.49\\
$\Delta \epsilon_s=2 MeV$\tablenotemark[1]& 1.5 & 0.5 & 3.50 & 1.39 & 2.21\\
\tableline
$^{22}Ne$ & 1 & 1 & 2.28 & 0.029~(0.36)\tablenotemark[2] & 1.81\\
$T=1~\rightarrow T=1$ & 1.5 & 1 & 3.98 & 0.046(0.44) & 3.20\\
                      & 1 & 0.5 & 2.51 & 0.030~(0.37) & 2.02\\
                      & 1.5 & 0.5 & 4.51 & 0.054~(0.46) & 3.69\\
$\Delta \epsilon_s=2 MeV$          & 1.5 & 0.5 & 5.79 & 0.068~(0.486)
& 4.32\\\\ 

$^{22}Ne$ & 1 & 1 & 0.34 & 0.27 & 0.16\\
$T=1~\rightarrow T=2$ & 1.5 & 1 & 0.38 & 0.31 & 0.39\\
          & 1 & 0.5 & 0.32 & 0.28 & 0.17\\
          & 1.5 & 0.5 & 0.41 & 0.32 & 0.51\\
$\Delta \epsilon_s=2 MeV$ & 1.5 & 0.5 & 0.48 & 0.38 & 0.76\\
\tableline
$^{24}Mg$ & 1 & 1 & 3.78 & 2.08 & 1.73\\
          & 1.5 & 1 & 4.95 & 1.75 & 3.35\\
          & 1   & 0.5 & 3.87 & 2.02 & 1.94\\
          & 1.5 & 0.5 & 5.43 & 1.69 & 4.04\\
$\Delta \epsilon_s=2 MeV$ & 1.5 & 0.5 & 6.45 & 1.63 & 5.46\\
\tableline
$^{28}Si$\tablenotemark[3] & 1 & 1 & 4.44 & 2.24 & 2.39 \\
          & 1.5 & 1 & 6.38 & 1.60 & 5.33\\
          & 1   & 0.5 & 4.59 & 2.06 & 2.79\\
          & 1.5 & 0.5 & 7.54 & 1.46 & 6.97\\
$\Delta \epsilon_s=2 MeV$ & 1.5 & 0.5 & 8.94 & 1.36 & 9.04 \\
\tableline
$^{32}S$ & 1 & 1 & 4.76 & 2.32 & 2.73 \\
         & 1.5 & 1 & 8.69 & 1.57 & 8.52\\
         & 1   & 0.5 & 5.43 & 2.17 & 3.76\\
         & 1.5 & 0.5 & 10.96 & 1.15 & 12.03\\
$\Delta \epsilon_s=2 MeV$ & 1.5 & 0.5 & 10.19 & 1.24 & 2.21\\
\tableline
$^{36}Ar$ & 1 & 1 & 3.35 & 1.44 & 2.25\\
          & 1.5 & 1 & 5.19 & 0.95 & 5.20\\
          & 1   & 0.5 & 3.72 & 1.33 & 2.87\\
          & 1.5 & 0.5 & 5.71 & 0.87 & 6.05\\
$\Delta \epsilon_s=2 MeV$ & 1.5 & 0.5 & 5.48 & 0.90 & 5.72\\
\tableline
$^{44}Ti$ & 1 & 1 & 2.41 & 1.50 & 0.89\\
          & 1.5 & 1 & 3.41 & 1.63 & 1.79\\
          & 1 & 0.5 & 2.42 & 1.47 & 0.94\\
          & 1.5 & 0.5 & 3.64 & 1.64 & 2.04\\
$\Delta \epsilon_p=2 MeV$\tablenotemark[4] & 1.5 & 0.5 & 5.56 & 2.13 &
3.53\\ 
\end{tabular}
\tablenotetext[1] {The $1s$ orbital is raised by 2 $MeV$ above the
$0d$ orbital}
\tablenotetext[2] {The first number in this column is the {\em
isoscalar} summed strength; the second number (in parentheses) is the
{\em isovector} summed strength.}
\tablenotetext[3] {For ($x,y$)=(1,1) the sum is over the 500 lowest
states, but for all the other ($x,y$) pairs the sum is over the lowest
50 states only.}
\tablenotetext[4] {The $1p$ orbital is raised by 2 $MeV$ above the
$0f$ orbital}
\end{table}

\begin{table}
\caption{Summed isovector $M1$ strength (First Ten States and All
States) for $x=1.5$, $y=0.5$} 
\begin{tabular}{llll}
$^{20}Ne$ & $B(M1)$ & $B_l$ & $B_\sigma$\\
1st 10 States & 2.05 & 0.90 & 1.22\\
All States     & 2.54 & 1.10 & 1.49\\
\tableline
$^{22}Ne~T=1~\rightarrow T=1$ & & & \\
1st 10 States & 3.71 & 0.23 & 2.88\\
All States     & 4.51 & 0.46 & 3.69\\\\

$^{22}Ne~T=1~\rightarrow T=2$ & & & \\
1st 10 States & 0.23 & 0.25 & 0.36\\
All States     & 0.41 & 0.31 & 0.51\\
\tableline
$^{24}Mg$ & & & \\
1st 10 States & 3.45 & 0.94 & 1.79\\
All States     & 5.43 & 1.68 & 4.03\\
\tableline
$^{28}Si$ & & & \\
1st 10 States & 5.30 & 0.66 & 4.76\\
All States    & 7.54 & 1.46 & 6.97\\
\tableline
$^{32}S$ & & & \\
1st 10 States & 10.13 & 0.53 & 11.16\\
All States     & 10.98 & 1.15 & 12.03\\
\tableline
$^{36}Ar$ & & & \\
1st 10 States & 5.50 & 0.63 & 5.95\\
All States     & 5.71 & 0.87 & 6.05\\
\tableline
$^{44}Ti$ & & & \\
1st 10 States & 2.73 & 1.18 & 1.10\\
All States     & 3.64 & 1.64 & 2.04\\
\end{tabular}
\end{table}

\begin{table}
\caption{Same as Table~III ($x=1.5,~y=0.5$) but with $\Delta
\epsilon_s=2 MeV$ for the $s-d$ shell nuclei and $\Delta \epsilon_p=2
MeV$ for $^{44}Ti$}
\begin{tabular}{llll}
$^{20}Ne$ & $B(M1)$ & $B_l$ & $B_\sigma$\\
1st 10 States & 3.39 & 1.25 & 2.09\\
All States     & 3.50 & 1.39 & 2.21\\
\tableline
$^{22}Ne~T=1~\rightarrow T=1$ & & & \\
1st 10 States & 5.20 & 0.25 & 4.18\\
All States     & 5.79 & 0.49 & 4.80\\\\

$^{22}Ne~T=1~\rightarrow T=2$ & & & \\
1st 10 States & 0.36 & 0.29 & 0.67\\
All States     & 0.48 & 0.38 & 0.76\\
\tableline
$^{24}Mg$ & & & \\
1st 10 States & 4.59 & 0.81 & 3.38\\
All States     & 6.45 & 1.63 & 5.45\\
\tableline
$^{28}Si$ & & & \\
1st 10 States & 7.08 & 0.52 & 7.24\\
All States     & 8.94 & 1.36 & 9.04\\
\tableline
$^{32}S$ & & & \\
1st 10 States & 8.81 & 0.41 & 9.45\\
All States     & 10.19 & 1.24 & 10.91\\
\tableline
$^{36}Ar$ & & & \\
1st 10 States & 5.10 & 0.48 & 5.21\\
All States     & 5.48 & 0.90 & 5.72\\
\tableline
$^{44}Ti$ & & & \\
1st 10 States & 3.73 & 1.32 & 1.26\\
All States     & 5.56 & 2.13 & 3.52\\
\end{tabular}
\end{table}

\begin{table}[htb]
\caption{$B_l$, $B_\sigma$ and $B(M1)$ strengths in $\mu_N^2$ for
states with $B(M1)$ (usually) $\ge 0.3 \mu_N^2$ as a function of $x$
and $y$ -the strengths of the spin-orbit and tensor interactions
respectively} 
\begin{tabular}{lcccccc}
&&&&&& \\
\hline
Nucleus & $x$ & $y$ & $E_x (MeV)$ & $B_l$ & $B_\sigma$ & $B(M1)$ \\
\hline
$^8Be$  & 1 & 1 & 13.72 & 0.257 & 0.115 & 0.716\\
        & 1.5 & 1 & 13.06 & 0.267 & 0.218 & 0.967\\
        & 1 & 0.5 & 13.75 & 0.255 & 0.139 & 0.770\\
        & 1.5 & 0.5 & 13.05 & 0.258 & 0.249 & 1.014\\
\hline
$^{10}Be$ & 1 & 1 & 7.68 & 0.017 & 1.512 & 1.849\\
          & 1.5 & 1 & 8.84 & 0.000 & 2.837 & 2.837\\
          & 1 & 0.5 & 7.71 & 0.001 & 2.333 & 1.966\\
          & 1.5 & 0.5 & 8.45 & 0.004 & 0.856 & 0.774\\
          &     &     & 9.26 & 0.000 & 2.232 & 2.695\\
\hline
$^{12}C$ & 1 & 1 & 13.60 & 0.031 & 0.583 & 0.886\\
         & 1.5 & 1 & 13.08 & 0.004 & 1.710 & 1.887\\
         & 1 & 0.5 & 13.33 & 0.012 & 0.901 & 1.121\\
         & 1.5 & 0.5 & 13.11 & 0.000 & 2.571 & 2.545\\
\hline
$^{20}Ne$ & 1 & 1 & 12.40 & 0.434 & 0.064 & 0.830\\
          &   &   & 19.21 & 0.136 & 0.047 & 0.344\\
          & 1.5 & 1 & 10.73 & 0.470 & 0.288 & 1.494\\
          & 1   & 0.5 & 12.18 & 0.240 & 0.003 & 0.187\\
          &     &     & 12.25 & 0.238 & 0.136 & 0.735\\
          & 1.5 & 0.5 & 10.66 & 0.475 & 0.346 & 1.631\\
$\Delta \epsilon_s=2 MeV$ & 1.5 & 0.5 & 9.757 & 0.511 & 0.623 & 2.263\\
             &     &     & 16.91 & 0.037 & 0.532 & 0.307\\
\hline
$^{22}Ne$ & 1 & 1 & 8.45 & 0.005 & 1.121 & 1.464\\
          & 1.5 & 1 & 7.06 & 0.176 & 0.053 & 0.461\\
          &     &   & 8.10 & 0.015 & 0.875 & 0.944\\
          &     &   & 10.00 & 0.006 & 0.370 & 0.374\\
          &     &   & 10.61 & 0.012 & 1.086 & 1.359\\
          &     &   &       &       &       &      \\
          & 1   & 0.5 & 8.63 & 0.001 & 1.492 & 1.800\\
          & 1.5 & 0.5 & 6.99 & 0.199 & 0.170 & 0.785\\
          &     &     & 8.57 & 0.011 & 1.077 & 1.065\\
          &     &     & 10.95 & 0.001 & 1.275 & 1.518\\
          &     &   &       &       &       &      \\
$\Delta \epsilon_s=2 MeV$ & 1.5 & 0.5 & 7.320 & 0.152 & 0.732 & 1.711\\
             &     &     & 8.575 & 0.009 & 1.664 & 1.544\\
             &     &     & 11.54 & 0.004 & 1.111 & 1.279\\
\hline
$^{24}Mg$ & 1 & 1 & 10.27 & 0.280 & 0.492 & 1.516\\
          & 1.5 & 1 & 9.35 & 0.204 & 0.943 & 2.024\\
          &     &   & 12.66 & 0.076 & 0.077 & 0.304\\
          & 1   & 0.5 & 10.17 & 0.280 & 0.595 & 1.693\\
          & 1.5 & 0.5 & 9.32  & 0.195 & 1.024 & 2.111\\
          &     &     & 9.466 & 0.083 & 0.265 & 0.644\\
          &     &   &       &       &       &      \\
$\Delta \epsilon_s=2 MeV$ & 1.5 & 0.5 & 9.401 & 0.183 & 2.316 & 3.799\\
             &     &     & 14.89 & 0.06  & 0.163 & 0.422\\
\hline
$^{28}Si$ & 1 & 1 & 10.31 & 0.055 & 0.860 & 1.352\\
          &   &   & 11.78 & 0.122 & 0.123 & 0.491\\
          &   &   & 12.97 & 0.359 & 0.001 & 0.387\\
          &   &   & 14.53 & 0.399 & 0.002 & 0.350\\
          &     &   &       &       &       &      \\
          & 1.5 & 1 & 9.64  & 0.075 & 2.810 & 3.801\\
          & 1 & 0.5 & 10.11 & 0.064 & 1.413 & 2.06\\
          &   &     & 14.54 & 0.383 & 0.001 & 0.352\\
          &     &   &       &       &       &      \\
          & 1.5 & 0.5 & 9.44 & 0.221 & 2.266 & 3.903\\
          &     &     & 10.54 & 0.062 & 1.313 & 0.804\\
          &     &   &       &       &       &      \\
$\Delta \epsilon_s=2 MeV$ & 1.5 & 0.5 & 10.03 & 0.007 & 4.258 &
4.612\\ 
                          &     &     & 10.99 & 0.005 & 0.486 & 
0.395\\
                          &     &     & 11.64 & 0.028 & 1.073 & 
0.756\\
                          &     &     & 12.71 & 0.136 & 0.252 & 
0.759\\
\hline
$^{32}S$ & 1 & 1 & 10.28 & 0.009 & 0.855 & 1.039\\
         &   &   & 11.17 & 0.111 & 0.266 & 0.721\\
         &   &   & 12.12 & 0.423 & 0.009 & 0.555\\
         &     &   &       &       &       &      \\
         & 1.5 & 1 & 9.63 & 0.007 & 4.574 & 4.212\\
         &     &   & 10.53 & 0.001 & 1.378 & 1.447\\
         &     &   & 11.77 & 0.163 & 0.242 & 0.801\\
         &     &   &       &       &       &      \\
         & 1   & 0.5 & 9.78 & 0.001 & 1.447 & 1.529\\
         &     &     & 11.14 & 0.097 & 0.753 & 1.391\\
         &     &   &       &       &       &      \\
         & 1.5 & 0.5 & 10.05 & 0.017 & 3.372 & 2.912\\
         &     &     & 11.21 & 0.018 & 5.976 & 5.341\\
         &     &     & 13.01 & 0.116 & 0.231 & 0.676\\
         &     &     & 13.81 & 0.105 & 0.979 & 0.442\\
         &     &   &       &       &       &      \\
\hline
$^{36}Ar$ & 1 & 1 & 10.70 & 0.002 & 1.180 & 1.290\\
          &   &   & 13.79 & 0.126 & 0.242 & 0.718\\
          &   &   & 15.50 & 0.237 & 0.005 & 0.311\\
          &     &   &       &       &       &      \\
          & 1.5 & 1 & 11.04 & 0.009 & 2.699 & 2.433\\
          &     &   & 12.10 & 0.053 & 0.105 & 0.307\\
          &     &   & 16.66 & 0.000 & 0.488 & 0.491\\
          &     &   &       &       &       &      \\
          & 1 & 0.5 & 10.97 & 0.016 & 1.639 & 1.981\\
          &   &     & 12.29 & 0.194 & 0.057 & 0.460\\
          &   &     & 16.41 & 0.071 & 0.174 & 0.466\\
          &     &   &       &       &       &      \\
          & 1.5 & 0.5 & 11.91 & 0.018 & 3.069 & 2.622\\
          &     &     & 12.97 & 0.041 & 1.010 & 1.457\\
          &     &     & 13.73 & 0.017 & 0.376 & 0.553\\
          &     &     & 17.65 & 0.002 & 0.431 & 0.488\\
          &     &   &       &       &       &      \\
$\Delta \epsilon_s=2 MeV$ & 1.5 & 0.5 & 10.78 & 0.073 & 0.817 & 0.402\\
             &     &     & 12.02 & 0.028 & 1.900 & 2.386\\
             &     &     & 12.30 & 0.025 & 0.519 & 0.771\\
             &     &     & 14.87 & 0.016 & 0.526 & 0.387\\
             &     &     & 15.16 & 0.027 & 0.509 & 0.300\\
             &     &     & 16.52 & 0.006 & 0.471 & 0.584\\
         &     &   &       &       &       &      \\
\hline
$^{44}Ti$    & 1   & 1   & 9.22  & 0.275 & 0.122 & 0.763\\
             &     &     & 11.91 & 0.527 & 0.015 & 0.365\\
         &     &   &       &       &       &      \\
             & 1.5 & 1   & 9.01  & 0.407 & 0.397 & 1.610\\
             &     &     & 9.50  & 0.356 & 0.007 & 0.465\\
         &     &   &       &       &       &      \\
             & 1   & 0.5 & 9.19  & 0.266 & 0.148 & 0.810\\
             &     &     & 12.76 & 0.020 & 0.205 & 0.352\\
         &     &   &       &       &       &      \\
             &     &     & 8.98  & 0.471 & 0.474 & 1.889\\
             &     &     & 9.49  & 0.303 & 0.001 & 0.340\\
\end{tabular}
\end{table}

\begin{table}[htb]
\caption{$B_l$, $B_\sigma$ and $B(M1)$ strengths in $\mu_N^2$ for
states of interest not contained in Table~X, and for $x=1.5~y=0.5$}

\begin{tabular}{lcccccc}
&&&&&& \\
\hline
Nucleus &  $E_x (MeV)$ & $B_l$ & $B_\sigma$ & $B(M1)$ \\
\hline
$^8Be$  & 17.42 & 0.230 & 0.185 & 0.003\\
        & 19.51 & 0.141 & 0.023 & 0.050\\
        &       &       &       &      \\
$^{10}Be$  & 16.11 & 0.020 & 0.000 & 0.025\\
 $T=1$     & 19.50 & 0.026 & 0.001 & 0.011\\
           & 20.57 & 0.039 & 0.010 & 0.009\\
   &       &       &       &      \\
 $T=2$     & 18.74 & 0.089 & 0.217 & 0.028\\
           & 23.47 & 0.046 & 0.012 & 0.010\\
   &       &       &       &      \\
$^{12}C$ & 17.46 & 0.284 & 0.074 & 0.068\\
         & 20.18 & 0.167 & 0.005 & 0.112\\
        &       &       &       &      \\
$^{20}Ne$ & 11.50 & 0.136 & 0.108 & 0.001\\
          & 18.09 & 0.239 & 0.361 & 0.013\\
$\Delta \epsilon_s=2 MeV$ & 12.80 & 0.290 & 0.162 & 0.018 \\
        &       &       &       &      \\
$^{24}Mg$ & 12.14 & 0.205 & 0.001 & 0.235\\
          & 13.92 & 0.237 & 0.384 & 0.017\\
          & 15.00 & 0.115 & 0.002 & 0.146\\
$\Delta \epsilon_s=2 MeV$ & 13.08 & 0.177 & 0.144 & 0.002\\
        &       &       &       &      \\
$^{28}Si$ & 13.26 & 0.142 & 0.381 & 0.058\\
$\Delta \epsilon_s=2 MeV$ & 10.79 & 0.104 & 0.003 & 0.140\\
                          & 13.25 & 0.196 & 0.682 & 0.147\\
                          & 14.47 & 0.003 & 0.214 & 0.168\\
          &       &       &       &      \\
$^{32}S$  & 17.01 & 0.140 & 0.021 & 0.270 \\
$\Delta \epsilon_s=2 MeV$ & 10.98 & 0.149 & 0.525 & 0.114\\
                          & 15.10 & 0.106 & 0.415 & 0.015\\
          &       &       &       &      \\
$^{36}Ar$ & 7.80  & 0.305 & 0.320 & 0.000\\
$\Delta \epsilon_s=2 MeV$ & 7.93  & 0.291 & 0.271 & 0.000\\
          &       &       &       &      \\
$^{44}Ti$ & 11.52 & 0.127 & 0.028 & 0.035\\
          & 12.77 & 0.095 & 0.469 & 0.142\\
$\Delta \epsilon_p=2 MeV$ & 9.93  & 0.323 & 0.130 & 0.043\\
\end{tabular}
\end{table}

\begin{table}
\caption{Isoscalar and Isovector $B(E2)$ (in $e^2fm^4$) as a function of the
strengths ($x,y$) of the spin-orbit and tensor interaction in small
space.}
\begin{tabular}{lllll}
Nucleus & $x$ & $y$ & $B(E2)_{isoscalar}^a$ & $B(E2)_{isovector}^b$ \\
\tableline
$^8Be$    & 1 & 1 & 71.27 & 9.26\\
          & 1.5 & 1 & 70.41 & 10.19\\
          & 1   & 0.5 & 71.24 & 9.35\\
          & 1.5 & 0.5 & 70.16 & 10.44\\
         &     &     &       &\\
$^{10}Be$ & 1 & 1 & 73.18 & 29.83\\
$T=1~\rightarrow T=1$ & 1.5 & 1 & 71.49 & 27.90\\
                      & 1   & 0.5 & 72.44 & 28.47 \\
                      & 1.5 & 0.5 & 70.29 & 27.38\\
         &     &     &       &\\            
$^{10}Be$ & 1 & 1 & 0 & 3.144 \\
$T=1~\rightarrow T=2$ & 1.5 & 1 & 0 & 3.512  \\          
                      & 1   & 0.5 & 0 & 3.164 \\
                      & 1.5 & 0.5 & 0 & 3.564 \\
         &     &     &       &\\            
$^{12}C$ & 1 & 1 & 82.73 & 13.56 \\
         & 1.5 & 1 & 78.43 & 17.72 \\
         & 1 & 0.5 & 82.00 & 14.16 \\
         & 1.5 & 0.5 & 75.30 & 19.20\\
         &     &     &       &\\
\tableline
$^{20}Ne$ & 1 & 1 & 315.1 & 53.44\\
          & 1.5 & 1 & 313.5 & 45.24\\
          & 1 & 0.5 & 316.0 & 52.68\\
          & 1.5 & 0.5 & 311.1 & 44.56\\
$\Delta \epsilon_s=2 MeV$ & 1.5 & 0.5 & 266.7 & 39.80\\
\tableline
$^{22}Ne$ & 1 & 1 &  352.7 & 108.3\\
$T=1~\rightarrow T=1$ & 1.5 & 1 & 350.4 & 101.1\\
                      & 1 & 0.5 & 353.8 & 107.9\\
                      & 1.5 & 0.5 & 344.8 & 98.64\\
$\Delta \epsilon_s=2 MeV$ & 1.5 & 0.5 & 298.0 & 90.17\\\\

$^{22}Ne$ & 1 & 1 & 0 & 22.12\\
$T=1~\rightarrow T=2$ & 1.5 & 1 & 0 & 16.39 \\
          & 1 & 0.5 & 0 & 21.26\\
          & 1.5 & 0.5 & 0 & 15.12\\
$\Delta \epsilon_s=2 MeV$ & 1.5 & 0.5 & 0 & 14.48\\
\tableline
$^{24}Mg$ & 1 & 1 & 515.8 & 71.07\\
          & 1.5 & 1 & 511.3 & 78.89 \\
          & 1   & 0.5 & 515.6 & 72.87\\
          & 1.5 & 0.5 & 508.9 & 80.68\\
$\Delta \epsilon_s=2 MeV$ & 1.5 & 0.5 & 481.8 & 82.86\\
\tableline
$^{28}Si^c$ & 1 & 1 & 651.8 & N/A\\
          & 1.5 & 1 & 574.0 & N/A \\
          & 1   & 0.5 & 635.4 & N/A\\
          & 1.5 & 0.5 & 546.9 & 26.4 \\
$\Delta \epsilon_s=2 MeV$ & 1.5 & 0.5 & 439.5 & N/A\\
\tableline
$^{32}S$ & 1 & 1 & 529.8 & 80.41\\
         & 1.5 & 1 & 337.2 & 123.8 \\
         & 1   & 0.5 & 488.5 & 89.66\\
         & 1.5 & 0.5 & 254.8 & 130.3\\
$\Delta \epsilon_s=2 MeV$ & 1.5 & 0.5 & 321.7 & 124.7\\
\tableline
$^{36}Ar$ & 1 & 1 & 331.2 & 59.01\\
          & 1.5 & 1 & 289.0 & 83.00 \\
          & 1   & 0.5 & 322.6 & 63.87\\
          & 1.5 & 0.5 & 280.9 & 85.71\\
$\Delta \epsilon_s=2 MeV$ & 1.5 & 0.5 & 307.9 & 75.91\\
\tableline
$^{44}Ti$ & 1 & 1 & 963.9 & 119.8\\
          & 1.5 & 1 & 953.7 & 109.9\\
          & 1 & 0.5 & 967.1 & 120.3\\
          & 1.5 & 0.5 & 946.9 & 109.2\\
$\Delta \epsilon_p=2 MeV$ & 1.5 & 0.5 &  & \\
\end{tabular}
(a) $e_p=1.5,~e_n=0.5$\\
(b) $e_p=1,~e_n=-1$\\
(c) For ($x,y$)=(1,1) the sum is over the 500 lowest states, but for
all the other ($x,y$) pairs the sum is over the lowest 50 states only. 
\end{table}

\end{small}

\begin{table}[htb]
\caption{B(M1) strength in $^{12}$C in 0 $\hbar \omega$ space (0
$\hbar \omega$).} 
\begin{tabular}{lcccc}
&&&& \\
\hline
& \multicolumn{2}{c}{$1^+;0$ state} & \multicolumn{2}{c}{$1^+;1$ state} \\
& $E_x$ & B(M1)\tablenotemark[1] & $E_x$ & B(M1) \\
& (MeV) & ($10^{-2} \mu^2_N$) & (MeV) & ($\mu^2_N$) \\
\hline
Experiment  & 12.71 & 4.0(3) & 15.11 & 2.63(8) \\ 
X=1,Y=1  & 11.80 & 0.26 & 13.60 & 0.89 \\
X=1.5,Y=1 & 11.03 & 0.86 & 13.08 & 1.89 \\
X=1,Y=0.5  & 12.30 & 0.54 & 13.33 & 1.12 \\
X=1.5,Y=0.5 & 11.80 & 1.65 & 13.11 & 2.55 \\
\end{tabular}
\tablenotetext[1] {It should be noted that $B(M1;0)$ is enhanced by
isospin mixing. One should compare the results of our calculations to
the Cohen-Kurath shell model calculation which found that 
$B(M1;0) = 1.4 \times 10^{-2}$ $\mu_N^2$, a value which agrees well
with our results, especially for $x=1.5~y=0.5$}
\end{table}

\begin{table}[htb]
\caption{B(E2) strength in $^{12}$C}
\begin{tabular}{lcccc}
&&&& \\
\hline
& \multicolumn{2}{c}{$2^+;0$ state} & \multicolumn{2}{c}{$2^+;1$ state} \\
& $E_x$ & $B(E2)$ & $E_x$ & $B(E2)$ \\
& (MeV) & e$^2$fm$^4$ & (MeV) & e$^2$fm$^4$ \\
($e_p,e_n$)& & (1.5,0.5) & & (1.5,0.5) \\ 
\hline
Expt  & 4.44 & 39.9(2.2) & 16.11 & 2.0(0.2) \\ 
X=1,Y=1     & 3.80 & 80.6  & 15.78 & 2.6 \\
X=1.5,Y=1   & 3.82 & 75.3 & 14.44 & 3.1 \\
X=1,Y=0.5   & 3.79 & 80.1 & 14.86 & 1.2  \\
            &      &      & 15.58 & 1.6  \\
X=1.5,Y=0.5 & 4.11 & 72.9 & 14.25 & 3.8  \\
\hline
\end{tabular}
\end{table}

\begin{table}[htb]
\caption{B(M1) strength in $^{20}$Ne.}
\begin{tabular}{lcccc}
&&&& \\
\hline
& \multicolumn{4}{c}{$1^+;1$ state} \\
& $E_x$ & $B(M1)$\tablenotemark[1] & $B_{\sigma}$ & $B_l$ \\
& (MeV) & ($\mu^2_N$) & ($\mu^2_N$) & ($\mu^2_N$) \\
\hline
Expt  & 11.26 & 2.02(36) & 0.49(6) & 0.52(15) \\ 
X=1,Y=1     & 12.40 & 0.83     & 0.06    & 0.43 \\
X=1.5,Y=1   & 10.73 & 1.49     & 0.29    & 0.47 \\
X=1,Y=0.5   & 12.19 & 0.19     & 0.03    & 0.24 \\
            & 12.25 & 0.73     & 0.14    & 0.24 \\
X=1.5,Y=0.5 & 10.61 & 1.63     & 0.35    & 0.47 \\
\hline
\end{tabular}
\tablenotetext[1] {It should be noted that only one $M1$ transition is
established, but $B_{\sigma}$ and $B_l$ can be separated from
comparison of (e,e$'$) and (p,p$'$). Constructive interference assumed.}
\end{table}

\begin{table}[htb]
\caption{B(E2) strength in $^{20}$Ne}
\begin{tabular}{lcc|cc}
&&&& \\
\hline
& \multicolumn{2}{c}{$2^+;0$ state} & \multicolumn{2}{c}{$2^+;0$ state} \\
& $E_x$ & $B(E2)$ & $E_x$ & $B(E2)$ \\
& (MeV) & (e$^2$fm$^4$)& (MeV) & (e$^2$fm$^4$) \\
($e_p,e_n$) & & (1.5,0.5) & & (1.5,0.5) \\
\hline
Expt  & 1.63 & 340(22) & 7.42 & $<$0.83 \\ 
            & 7.83 & 12.0(1.5) & 10.27 & 1.4(4) \\ 
\hline
X=1,Y=1     & 2.92 & 299 & 9.21 & 2.9 \\
            & 11.82 & 4.4 & 11.12 & 4.1 \\
            &       &     & 11.76 & 0.97\\
\hline
X=1.5,Y=1   & 2.40 & 298 & 7.82 & 0.95 \\
            & 10.14 & 0.7 &  &  \\
            & 10.75 & 3.6 & 9.90 & 1.3 \\
            & 11.08 & 0.4 & 10.76 & 3.0\\
\hline
X=1,Y=0.5   & 2.84 & 301 & 9.01 & 2.6 \\
            & 11.08 & 0.4 & 11.20 & 3.9 \\
            & 11.74 & 3.8 & 11.65 & 1.1 \\
\hline
X=1.5,Y=0.5 & 2.37 & 296 & 7.61 & 0.73 \\
            & 10.05 & 1.8 & 9.81 & 1.7 \\
            & 10.73 & 3.0 & 10.90 & 2.5 \\
\hline
\end{tabular}
\end{table}

\begin{table}[htb]
\caption{B(M1) strength in $^{22}$Ne}
\begin{tabular}{lcccccc}
&&&&&& \\
\hline
& \multicolumn{2}{c}{$1^+;1$ state} & \multicolumn{2}{c}{$1^+;1$ state} 
& \multicolumn{2}{c}{$1^+;1$ state} \\
& $E_x$ & $B(M1)$ & $E_x$ & $B(M1)$ & $E_x$ & $B(M1)$ \\
& (MeV) & ($\mu_N^2$) & (MeV) & ($\mu_N^2$) & (MeV) & ($\mu_N^2$) \\
\hline
Expt.   & 5.33 & 0.43(17) & 6.85 & 1.36(56) & 9.18 & 1.80(7) \\ 
($x$,$y$) &    &          &      &          &      & \\
(1,1)     & 5.70 & 0.02 & 6.85 & 0.14 & 8.45 & 1.46 \\
(1.5,1)   & 7.06 & 0.46 & 8.10 & 0.94 & 10.61 & 1.36 \\
(1,0.5)   & 5.59 & 0.007& 7.76 & 0.15 & 8.63 & 1.80 \\
(1.5,0.5) & 6.99 & 0.79 & 8.56 & 1.07 & 10.95 & 1.52 \\
\hline
\end{tabular}
\end{table}

\begin{table}[htb]
\caption{B(E2) strength in $^{22}$Ne.}
\begin{tabular}{lcc|cc}
&&&& \\
\hline
& \multicolumn{2}{c}{$2^+;1$ state} & \multicolumn{2}{c}{$2^+;1$
state} \\ 
& $E_x$ & $B(E2)$ & $E_x$ & $B(E2)$ \\ 
& (MeV) & (e$^2$fm$^4$) & (MeV) & (e$^2$fm$^4$) \\
($e_p,e_n$)\tablenotemark[1] & (1.5,0.5) & (1.5,0.5)\\ 
\hline
Expt  & 1.27 & 235(3) & 4.46 & $>4.6$ \\
            & 6.12 & 3.5(1.5) & 7.64 & 19(8) \\              
\hline
X=1,Y=1     & 2.27 & 199  & 3.10 & 113 \\
            & 4.21 & 2.9  & 5.85 & 12.4 \\
\hline
X=1.5,Y=1   & 1.98 & 289  & 2.77 & 7.5  \\
            & 4.27 & 8.7  & 5.03 & 22.9  \\
\hline
X=1,Y=0.5   & 2.28 & 243  & 3.02 & 67 \\
            & 4.18 & 6.5  & 5.72 & 13.0 \\
\hline
X=1.5,Y=0.5 & 1.90 & 290  & 3.00 & 0.004 \\
            & 4.43 & 7.1  & 4.88 & 26.1 \\
\hline
\end{tabular}
\tablenotetext[1] {Results for the other ($e_p$,$e_n$) combinations are
significantly worse. It should be noted that there is no experimental
evidence for isovector $E2$ transitions.} 
\end{table}

\begin{table}
\caption{Experimental $B(M1)$ values$^a$ for $^{24}Mg$}
\begin{tabular}{lc}
 $E_x$ & $B(M1)\uparrow$ \\
 $MeV$ & ($\mu^2_N$) \\
\hline
8.865  & 0.042 $\pm$ 0.011\\
9.820  & 0.260 $\pm$ 0.040\\
9.962  & 1.160 $\pm$ 0.150\\
10.711 & 3.180 $\pm$ 0.300\\
10.918 & 0.100 $\pm$ 0.020\\
11.002 & 0.048 $\pm$ 0.011\\
11.373 & 0.056 $\pm$ 0.013\\
11 851 & 0.033 $\pm$ 0.008\\
11.960 & 0.046 $\pm$ 0.010\\
12.145 & 0.035 $\pm$ 0.012\\
12.264 & 0.035 $\pm$ 0.009\\
12.406 & 0.047 $\pm$ 0.010\\
12.527 & 0.065 $\pm$ 0.011\\
12.807 & 0.179 $\pm$ 0.018\\
12.961 & 0.151 $\pm$ 0.012\\
13.081 & 0.024 $\pm$ 0.011\\
13.222 & 0.067 $\pm$ 0.009\\
13.675 & 0.030 $\pm$ 0.008\\
13.940 & 0.201 $\pm$ 0.014\\
14.094 & 0.026 $\pm$ 0.008\\
14.267 & 0.059 $\pm$ 0.009\\
\end{tabular}
(a) obtained with the use of the Wildenthal-Brown form factor.
\end{table}

\begin{table}
\caption[]{Experimental Excitation Energies and $B_\sigma$ \\
values$^a$ for $1^+~T=1$ states $^{24}Mg$}
\begin{tabular}{lc}
 $E_x$ & $B_\sigma$ \\
 $MeV$ & ($\mu^2_N$) \\
\hline
9.83 ($T=0$ ?) &  0.29 $\pm$ 0.02\\
9.97           &  0.38 $\pm$ 0.03\\
10.72          &  2.75 $\pm$ 0.20\\
12.53          &  0.60 $\pm$ 0.03\\
12.82          &  0.85 $\pm$ 0.03\\
12.96          &  0.37 $\pm$ 0.02\\
13.90          &  0.56 $\pm$ 0.09\\
16.12          &  0.31 $\pm$ 0.06\\
\end{tabular}
(a) Taken from ($p,p'$) data by Crawley $et.al.$~\cite{craw}. 
\end{table}

\begin{table}
\caption{Calculated $B(M1)\uparrow$ strengths to $J=1^+~T=1$ states in
$^{24}Mg$ obtained with the Realistic Interaction ($x=1.5~y=0.5$).}
\begin{tabular}{lccc}
 $E_x$ & $B(M1)_{physical}$ & $B_l$ & $B_\sigma$\\
 $MeV$ & ($\mu^2_N$) & ($\mu^2_N$) & ($\mu^2_N$) \\
\hline
9.33  & 2.11 & 0.19 & 1.02\\
9.47  & 0.64 & 0.08 & 0.27\\
10.50 & 0.03 & 0.00 & 0.03\\
11.91 & 0.19 & 0.07 & 0.03\\
12.14 & 0.23 & 0.21 & 0.001\\
13.49 & 0.03 & 0.01 & 0.008\\
13.92 & 0.02 & 0.24 & 0.38\\
14.24 & 0.02 & 0.02 & 0.00\\
14.59 & 0.03 & 0.00 & 0.05\\
15.00 & 0.15 & 0.12 & 0.002\\
\hline
Sum(10)$^a$ & 3.45 & 0.94 & 1.79\\
Total Sum$^b$ & 5.43 & 1.69 & 4.04\\
\end{tabular}
(a) Sum to the lowest 10 states shown in the table.\\
(b) Sum to all $J=1^+~T=1$ states in the model space.
\end{table}

\begin{table}
\caption{Calculated $B(M1)\uparrow$ strengths to $J=1^+~T=1$ states in
$^{24}Mg$ obtained with ($x=1.5~y=0.5$) and $\Delta \epsilon_s=2~MeV$.}
\begin{tabular}{lccc}
 $E_x$ & $B(M1)_{physical}$ & $B_l$ & $B_\sigma$\\
 $MeV$ & ($\mu^2_N$) & ($\mu^2_N$) & ($\mu^2_N$) \\
\hline
9.40  & 3.80  & 0.18 & 2.32\\
9.88  & 0.00  & 0.05  & 0.10\\
11.08 & 0.08  & 0.04  & 0.01\\
12.08 & 0.00  & 0.09  & 0.14\\
13.08 & 0.00  & 0.18  & 0.14\\
13.33 & 0.06  & 0.06  & 0.00\\
13.98 & 0.01  & 0.06  & 0.13\\
14.06 & 0.10  & 0.07  & 0.35\\
14.89 & 0.42  & 0.06  & 0.16\\
15.19 & 0.10  & 0.02  & 0.03\\
\hline
Sum(10) & 4.58 & 0.81 & 3.38\\
Total Sum & 6.45 & 1.63 & 5.46\\
\end{tabular}
\end{table}

\begin{table}[htb]
\caption{B(E2) strength in $^{24}$Mg}
\begin{tabular}{lcc|cc}
\hline
& \multicolumn{2}{c|}{$2^+;0$ state} & \multicolumn{2}{c}{$2^+;0$
state} \\
\hline
& $E_x$ & $B(E2)$ & $E_x$ & $B(E2)$ \\
& (MeV) & (e$^2$fm$^4$)& (MeV) & (e$^2$fm$^4$) \\
($e_p,e_n$) & & (1.5,0.5) & & (1.5,0.5) \\
\hline
Expt  & 1.369 & 435(9) & 4.238 & 33.3(2.4) \\ 
      & 7.349 & 11.8(4.7) & 9.004 & 3.3(0.6)\\ 
            &       &     &  & \\
\hline
X=1,Y=1     & 2.78 & 488 & 4.33 & 11.9 \\
            & 9.32 & 0.6 & 9.93 & 3.8\\
            &       &     &  & \\
\hline
X=1.5,Y=1   & 2.39 & 474 & 3.74 & 12.8 \\
            & 7.87 & 0.8 & 9.67 & 3.5 \\
            &  &  &  & \\
\hline
X=1,Y=0.5   & 2.71 & 486 & 4.24 & 13.4 \\
            & 9.07 & 0.8 & 10.0 & 2.7 \\
            &  &  &  &  \\
\hline
X=1.5,Y=0.5 & 2.33 & 479 & 3.62 & 4.9 \\
            & 7.44 & 1.51 & 9.31 & 4.4 \\
\end{tabular}
\end{table}
 
\begin{table}
\caption{Experimental $B(M1)$ values$^a$ for $^{28}Si$}
\begin{tabular}{lc}
 $E_x$ & $B(M1)\uparrow$ \\
 $MeV$ & ($\mu^2_N$) \\
\hline
10.594  & 0.19 $\pm$ 0.04\\
10.725  & 0.11 $\pm$ 0.01\\
10.901  & 0.90 $\pm$ 0.02\\
11.445  & 4.42 $\pm$ 0.20\\
12.331  & 0.87 $\pm$ 0.06\\
14.030  & 0.37 $\pm$ 0.02\\
15.147  & 0.23 $\pm$ 0.02\\
15.500  & 0.26 $\pm$ 0.03\\
17.56   & 0.18 $\pm$ 0.03\\
\end{tabular}
(a) Obtained through DWBA analysis.
\end{table}

\begin{table}
\caption[]{Experimental Excitation Energies and $B_\sigma$ \\
values$^a$ for $1^+~T=1$ states $^{28}Si$}
\begin{tabular}{lc}
 $E_x$ & $B_\sigma$ \\
 $MeV$ & ($\mu^2_N$) \\
\hline
9.72  & 0.39 $\pm$ 0.06\\
10.59 & 0.83 $\pm$ 0.83\\
10.73 & 0.32 $\pm$ 0.04\\
10.82 & 0.21 $\pm$ 0.04\\
10.90 & 0.35 $\pm$ 0.05\\
11.16 & 0.31 $\pm$ 0.07\\
11.45 & 3.32 $\pm$ 0.24\\
12.33 & 0.73 $\pm$ 0.14\\
12.99 & 0.23 $\pm$ 0.05\\
13.35 & 0.81 $\pm$ 0.14\\
14.03 & 1.31 $\pm$ 0.12\\
15.15 & 0.42 $\pm$ 0.04\\
15.50 & 0.12 $\pm$ 0.08\\
15.80 & 0.22 $\pm$ 0.02\\
Sum   & 9.51\\
\end{tabular}
(a) Taken from ($p,p'$) data by Crawley $et.al.$~\cite{craw}.
\end{table}

\begin{table}
\caption[]{Calculated $B(M1)\uparrow$ strengths to \\
$J=1^+~T=1$ states in $^{28}Si$ obtained with the Realistic \\
Interaction ($x=1.5~y=0.5$).}
\begin{tabular}{lccc}
 $E_x$ & $B(M1)_{physical}$ & $B_l$ & $B_\sigma$\\
 $MeV$ & ($\mu^2_N$) & ($\mu^2_N$) & ($\mu^2_N$) \\
\hline
9.43  & 3.90 & 0.22 & 2.27\\
10.54 & 0.80 & 0.06 & 1.31\\
11.05 & 0.00 & 0.03 & 0.02\\
12.07 & 0.02 & 0.03 & 0.00\\
12.65 & 0.20 & 0.03 & 0.41\\
13.27 & 0.06 & 0.14 & 0.38\\
13.32 & 1.39 & 0.06 & 0.02\\
14.11 & 0.02 & 0.00 & 0.02\\
14.61 & 0.11 & 0.06 & 0.33\\
14.78 & 0.05 & 0.02 & 0.01\\
\hline
Sum(10)$^a$ & 5.30 & 0.66 & 4.77\\
Total Sum$^b$ & 7.54 & 1.46 & 6.97\\
\end{tabular}
(a) Sum to the lowest 10 states shown in the table.\\
(b) Sum to all $J=1^+~T=1$ states in the model space.
\end{table}

\begin{table}
\caption[]{Calculated $B(M1)\uparrow$ strengths to \\
$J=1^+~T=1$ states in $^{28}Si$ obtained with ($x=1.5~y=0.5$)\\
and $\Delta \epsilon_s=2~MeV$.}
\begin{tabular}{lccc}
 $E_x$ & $B(M1)_{physical}$ & $B_l$ & $B_\sigma$\\
 $MeV$ & ($\mu^2_N$) & ($\mu^2_N$) & ($\mu^2_N$) \\
\hline
10.03  & 4.61 & 0.01 & 4.26\\
10.79  & 0.14 & 0.10 & 0.00\\
10.99  & 0.39 & 0.00 & 0.49\\
11.64  & 0.76 & 0.03 & 1.07\\
12.71  & 0.76 & 0.14 & 0.25\\
12.82  & 0.01 & 0.00 & 0.01\\
13.25  & 0.15 & 0.20 & 0.68\\
13.64  & 0.00 & 0.00 & 0.01\\
14.08  & 0.08 & 0.04 & 0.25\\
14.47  & 0.17 & 0.00 & 0.21\\
\hline
Sum(10) & 7.08 & 0.52 & 7.24\\
Total Sum & 8.94 & 1.36 & 9.04\\
\end{tabular}
\end{table}

\begin{table}[htb]
\caption[]{Experimental $B(E2)$ ($0_1^+ \rightarrow 2_1^+ T$) in
$^{28}$Si} 
\begin{tabular}{lcc}
 $E_x$ & $B(E2)$ & $T$\\
 ($MeV$) & ($e^2 fm^4$) & \\
1.779 & 334(6) & 0\\
7.381 & 8.5(0.6) & 0\\
7.416 & 4.5(0.5) & 0\\
7.993 & 7.7(1.1) & 0\\
8.259 & 0.73(0.17) & 0\\
9.381 & 1.3(0.5) & 1\\
\hline
\end{tabular}
\end{table}

\begin{table}[htb]
\caption[]{Theoretical $B(E2)$ ($0_1^+ \rightarrow 2_1^+ T$) in
$^{28}$Si calculated with $x=1.5~y=0.5$ ($e_p=1.5~e_n=0.5$)} 
\begin{tabular}{lcc}
 $E_x$ & $B(E2)$ & $T$\\
 ($MeV$) & ($e^2 fm^4$) & \\
2.41 & 496.7 & 0\\
4.64 & 4.3 & 0\\
6.87 & 1.4 & 0\\
8.00 & 26.7 & 0\\
8.59 & 8.8 & 1\\
9.34 & 0.3 & 0\\
9.92 & 0.7 & 0\\
\hline
\end{tabular}
\end{table}

\begin{table}
\caption[]{Experimental Excitation Energies and $B_\sigma$ \\
and $B(M1)$ values$^a$ for $1^+~T=1$ states $^{32}S$}
\begin{tabular}{lcc}
 $E_x$ & $B_\sigma$ & $B(M1)^b$ \\
 $MeV$ & ($\mu^2_N$) & ($\mu^2_N$) \\
\hline
7.63 & 0.42 $\pm$ 0.07 &\\
7.92 & 0.10 $\pm$ 0.02 &\\
8.13 & 1.46 $\pm$ 0.19 & 1.14 $\pm$ 0.18\\
9.66 & 0.13 $\pm$ 0.02 & 0.69 $\pm$ 0.18\\
11.13 & 4.08 $\pm$ 0.53 & 2.40 $\pm$ 0.22\\
11.63 & 2.38 $\pm$ 0.35 & 1.26 $\pm$ 0.20\\
11.88 & 0.37 $\pm$ 0.06 &\\
12.56 & 0.33 $\pm$ 0.06 &\\
13.90 & 0.24 $\pm$ 0.03 &\\
14.88 & 0.20 $\pm$ 0.04 &\\
15.58 & 0.28 $\pm$ 0.05 &\\
15.70 & 0.16 $\pm$ 0.04 &\\
15.84 & 0.26 $\pm$ 0.06 &\\
\end{tabular}
(a) Taken from ($p,p'$) data by Crawley $et.al.$~\cite{craw}.\\
(b) From ($e,e'$) data cited by Crawley $et.al.$~\cite{craw}.
\end{table}

\begin{table}
\caption[]{Experimental Excitation Energies and \\
$B(M1)$ values$^a$ for $1^+~T=1$ states $^{32}S$}
\begin{tabular}{lc}
 $E_x$  & $B(M1)$ \\
 $MeV$ & ($\mu^2_N$) \\
\hline
9.66 & 0.55 $\pm$ 0.24\\
9.98 & 0.09 $\pm$ 0.04\\
10.45 & 0.10 $\pm$ 0.08\\
10.90$^b$ & 0.33 $\pm$ 0.04\\
11.16 & 1.24 $\pm$ 0.13\\
11.50 & 0.10 $\pm$ 0.04\\
11.65 & 0.77 $\pm$ 0.14\\
12.19 & 0.14 $\pm$ 0.09\\
12.65 & 0.11 $\pm$ 0.04\\
12.98 & 0.07 $\pm$ 0.04\\
13.41$^b$ & 0.54 $\pm$ 0.06 \\
13.78 & 0.39 $\pm$ 0.05\\
13.97$^b$ & 0.20 $\pm$ 0.11\\
14.45 & 0.18 $\pm$ 0.05\\
\end{tabular}
(a) Taken from paper by Petraitis {\em et. al.}~\cite{petr}. \\
(b) May be $M1$ or $M2$.
\end{table}

\begin{table}
\caption[]{Calculated $B(M1)\uparrow$ strengths to $J=1^+~T=1$ states in
$^{32}S$ obtained with the Realistic Interaction ($x=1.5~y=0.5$).}
\begin{tabular}{lccc}
 $E_x$ & $B(M1)_{physical}$ & $B_l$ & $B_\sigma$\\
 $MeV$ & ($\mu^2_N$) & ($\mu^2_N$) & ($\mu^2_N$) \\
\hline
10.05 & 2.91 & 0.02 & 3.37\\
11.21 & 5.34 & 0.02 & 5.98\\
13.02 & 0.68 & 0.12 & 0.23\\
13.85 & 0.44 & 0.11 & 0.98\\
15.10 & 0.04 & 0.06 & 0.22\\
15.62 & 0.13 & 0.02 & 0.05\\
15.93 & 0.01 & 0.02 & 0.01\\
16.55 & 0.16 & 0.02 & 0.07\\
17.01 & 0.27 & 0.14 & 0.02\\
17.31 & 0.14 & 0.01 & 0.23\\
\hline
Sum(10) & 10.13 & 0.53 & 11.16\\
Total Sum & 10.98 & 11.52 & 12.03\\
\end{tabular}
\end{table}

\begin{table}
\caption[]{Calculated $B(M1)\uparrow$ strengths to $J=1^+~T=1$ states in
$^{32}S$ obtained with  ($x=1.5~y=0.5$) and $\Delta \epsilon_s=2~MeV$.}
\begin{tabular}{lccc}
 $E_x$ & $B(M1)_{physical}$ & $B_l$ & $B_\sigma$\\
 $MeV$ & ($\mu^2_N$) & ($\mu^2_N$) & ($\mu^2_N$) \\
\hline
 8.58 & 0.40 & 0.01 & 0.53\\
10.25 & 4.33 & 0.00 & 4.20\\
10.98 & 0.11 & 0.15 & 0.52\\
12.08 & 2.64 & 0.01 & 3.02\\
12.40 & 0.05 & 0.01 & 0.01\\
13.04 & 0.18 & 0.00 & 0.15\\
14.22 & 0.02 & 0.06 & 0.15\\
14.83 & 0.12 & 0.01 & 0.22\\
15.10 & 0.02 & 0.11 & 0.04\\
15.17 & 0.95 & 0.04 & 0.59\\
\hline
Sum(10) & 8.81 & 0.41 & 9.45\\
Total Sum & 10.19 & 1.24 & 10.91\\
\end{tabular}
\end{table}

\begin{table}[htb]
\caption[]{B(E2) strength in $^{32}$S}
\begin{tabular}{lcccc}
&&&& \\
\hline
& \multicolumn{2}{c}{$2^+;0$ state} & \multicolumn{2}{c}{$2^+;0$ state} \\
& $E_x$ & $B(E2)$ & $E_x$ & $B(E2)$ \\
& (MeV) & (e$^2$fm$^4$)& (MeV) & (e$^2$fm$^4$) \\
($e_p,e_n$) & & (1.5,0.5) & & (1.5,0.5) \\
\hline
Expt  & 2.23 & 306(9) & 4.49 & 58.1(6.9)\\ 
      & 5.55 & 3.9(1.0) & 7.12 & > 1.0\\
      &      &          &      &\\
X=1,Y=1     & 3.00 & 505 & 4.63 & 10.1\\
            & 9.9  & 0.8 &  11.3 & 1.6 \\
      &      &          &      &\\
X=1.5,Y=1   & 4.45 & 28.3 & 5.39 & 41.1\\
            & 9.4 & 0.6 & 10.4 & 12.2\\
      &      &          &      &\\
X=1,Y=0.5 & 3.19 & 460 & 4.62 & 16.2\\
          & 9.5 & 0.7 & 10.7 & 2.8\\
      &      &          &      &\\
X=1.5,Y=0.5 & 5.77 & 186 & 6.32 & 58.1\\
            & 12.6 & 3.9 & 11.3 & 27.3\\
\hline
\end{tabular}
\end{table}

\begin{table}[htb]
\caption{B(M1) strength in $^{36}$Ar. 
Only summed strength is discussed because of the large error bars 
of the (e,e$'$) experiment.}
\begin{tabular}{lc}
\\
\hline
& B(M1) \\
& ($\mu^2_N$) \\
\hline
Expt                  & 2.65 $\pm$ 0.1\\
X=1,Y=1               & 2.30 \\ 
X=1.5,Y=1             & 4.09 \\ 
X=1,Y=0.5             & 2.68 \\ 
X=1.5,Y=0.5           & 4.63 \\ 
\hline
\end{tabular}
\end{table}

\begin{table}[htb]
\caption{B(E2) strength in $^{36}$Ar}
\begin{tabular}{lcccc}
&&&& \\
\hline
& \multicolumn{2}{c}{$2^+;0$ state} & \multicolumn{2}{c}{$2^+;0$ state} \\
& $E_x$ & $B(E2)$ & $E_x$ & $B(E2)$ \\
& (MeV) & (e$^2$fm$^4$)& (MeV) & (e$^2$fm$^4$) \\
($e_p,e_n$) & & (1.5,0.5) & & (1.5,0.5) \\
\hline
Expt  & 1.97 & 340(40) & 4.44 & 14.0(2.5) \\ 
X=1,Y=1     & 2.85 & 312 & 7.10 & 2.7 \\
X=1.5,Y=1   & 2.23 & 247 & 5.96 & 6.3 \\
X=1,Y=0.5   & 2.77 & 301 & 6.96 & 5.0 \\
X=1.5,Y=0.5 & 2.19 & 236 & 5.88 & 3.6 \\
\hline
\end{tabular}
\end{table}

\begin{table}[htb]
\caption{B(E2) strength in $^{36}$Ar continued}
\begin{tabular}{lcccc}
&&&& \\
\hline
& \multicolumn{2}{c}{$2^+;0$ state} & \multicolumn{2}{c}{$2^+;1$ state} \\
& $E_x$ & $B(E2)$ & $E_x$ & $B(E2)$ \\
& (MeV) & (e$^2$fm$^4$)& (MeV) & (e$^2$fm$^4$) \\
($e_p,e_n$) & & (1.5,0.5) & & (1.5,0.5) \\
\hline
Expt  & 4.95 & $<21.4$ & 6.61 & 2.4(0.9) \\ 
X=1,Y=1     & 9.39 & 2.2 & 8.56 & 3.6 \\
X=1.5,Y=1   & 8.19 & 20.8 & 6.34 & 8.0 \\
X=1,Y=0.5   & 8.81 & 5.4 & 8.19 & 4.6 \\
X=1.5,Y=0.5 & 8.51 & 30.0 & 6.28 & 8.5 \\
\hline
\end{tabular}
\end{table}

\begin{table}[htb]
\caption{$B(E2)$ strength in $^{44}$Ti}
\begin{tabular}{lcccc}
&&&& \\
\hline
& \multicolumn{2}{c}{$2^+;0$ state} & \multicolumn{2}{c}{$2^+;0$ state} \\
& $E_x$ & $B(E2)$ & $E_x$ & $B(E2)$ \\
& (MeV) & (e$^2$fm$^4$) & (MeV) & (e$^2$fm$^4$) \\
\hline
Expt  & 1.08 & 540(140) & 2.53 & 14.2(2.4) \\ 
X=1,Y=1$^a$     & 2.08 & 924 &  6.12 & 5.0  \\
X=1.5,Y=1   & 1.67 & 913 &  5.65 & 2.1  \\
X=1,Y=0.5   & 2.02 & 929 &  6.06 & 4.0  \\
X=1.5,Y=0.5 & 1.65 & 908 &  5.68 & 2.3  \\
\hline
\end{tabular}
(a) The calculations were done with $e_p=1.5,~e_n=0.5$.
\end{table}

\end{document}